\documentclass{article}
\usepackage[utf8]{inputenc}
\usepackage{graphicx}
\usepackage{dcolumn}
\usepackage{amsmath}
\usepackage{amsfonts}
\usepackage{bm}% bold math
\usepackage{url}
\usepackage{enumitem}
\newcommand{\lp}{\left(}
\newcommand{\rp}{\right)}
\title{Physics Successfully Implements Lagrange Multiplier Optimization}
\author{Sri Krishna Vadlamani$^1$, Tianyao Patrick Xiao$^{1}$\footnote{Now at Sandia National Laboratories, Albuquerque, NM, USA} , Eli Yablonovitch$^1$}
\date{%
    $^1$Department of Electrical Engineering and Computer Sciences, University of California, Berkeley, CA, USA\\%
    \today
}

\begin{document}

\maketitle

\begin{abstract}
Optimization is a major part of human effort. While being mathematical, optimization is also built into physics. For example, physics has the principle of Least Action, the principle of Minimum Entropy Generation, and the Variational Principle. Physics also has physical annealing which, of course, preceded computational Simulated Annealing. Physics has the Adiabatic Principle which in its quantum form is called Quantum Annealing. Thus, physical machines can solve the mathematical problem of optimization, including constraints.

Binary constraints can be built into the physical optimization. In that case, the machines are digital in the same sense that a flip-flop is digital. A wide variety of machines have had recent success at optimizing the Ising magnetic energy. We demonstrate in this paper that almost all those machines perform optimization according to the Principle of Minimum Entropy Generation as put forth by Onsager. Further, we show that this optimization is in fact equivalent to Lagrange multiplier optimization for constrained problems. We find that the physical gain coefficients which drive those systems actually play the role of the corresponding Lagrange Multipliers.
\end{abstract}

\section{Introduction}
\label{intro}
Optimization is ubiquitous in today's world. Everyday applications of optimization range from aerodynamic design of vehicles by fluid mechanics and physical stress optimization of bridges in civil engineering to scheduling of airline crews and routing of delivery trucks in operations research. Furthermore, optimization is also indispensable in machine learning, reinforcement learning, computer vision, and speech processing. Given the preponderance of massive datasets and computations today, there has been a surge of activity in the design of hardware accelerators for neural network training and inference \cite{shen_deep_2017}.

We ask whether Physics can address optimization? There are a number of promising physical principles that drive dynamical systems toward an extremum. These are: the principle of Least Action, the principle of Minimum Entropy Generation, and the Variational Principle. Physics also has actual physical annealing
which preceded computational Simulated Annealing. Further, Physics has the Adiabatic Principle which in its quantum form is called Quantum Annealing.

In due course, we may learn how to use each of these principles to perform optimization. Let us consider the principle of minimum entropy generation in dissipative physical systems, such as resistive electrical circuits. It was shown by Onsager \cite{onsager_reciprocal_1931} that the equations of linear systems, like resistor networks, can be re-expressed as the minimization principle of a function $f(i_1,i_2,\dots,i_n)$ for currents $i_n$ in various branches of the resistor network. For a resistor network, the function $f$ contains the power dissipation, or entropy generation. By re-expressing a merit function in terms of power dissipation, the circuit itself will find the minimum of the merit function, or minimum power dissipation. Optimization is generally accompanied by constraints. For example, perhaps the constraint is that the final answers must be restricted to be $\pm 1$. Such a digitally constrained optimization produces answers compatible with any digital computer. 

There has been a series of machines created in the physics and engineering community to devise physics-based engines for the Ising problem. The Ising challenge is to find the minimum energy configuration of a large set of magnets. It's a very hard problem even when the magnets are restricted to only two orientations, North pole up or down \cite{lucas_ising_2013}. Our main insight in this paper is that most of these Ising solvers use hardware based on the Principle of Minimum Entropy generation. The natural evolution of these machines is toward a good, low-power-dissipation final state. Further, almost all of them implement the well-known Lagrange Multipliers method for constrained optimization.

An early work was by Yamamoto et al. in \cite{utsunomiya_mapping_2011} and this was followed by further work from their group \cite{wang_coherent_2013}, \cite{marandi_network_2014}, \cite{haribara_computational_2016}, \cite{mcmahon_fully_2016}, \cite{inagaki_large-scale_2016}, \cite{inagaki_coherent_2016}, \cite{leleu_destabilization_2019} and other groups \cite{xiao_optoelectronics_2019}, \cite{babaeian_single_2019}, \cite{mcquillan_oim_2019}, \cite{kalinin_global_2018}, \cite{roques-carmes_heuristic_2020}, \cite{goto_combinatorial_2019}, \cite{molnar_continuous-time_2018}, \cite{pierangeli_large-scale_2019}, \cite{pierangeli_scalable_2020}. These entropy generating machines range from coupled optical parametric oscillators, to RLC electrical circuits, to coupled exciton-polaritons, and silicon photonic coupled arrays. These types of machines have the advantage that they solve digital problems in the analog domain, which can be orders-of-magnitude faster and more energy-efficient than conventional digital chips that are limited by latency and the energy cost \cite{leleu_destabilization_2019}.

Within the framework of these dissipative machines, constraints can be readily included. In effect, these machines perform constrained optimization equivalent to the technique of Lagrange multipliers. We illustrate this connection by surveying 7 published physically distinct machines and showing that each minimizes entropy generation in its own way, subject to constraints; in fact, they perform Lagrange multiplier optimization. We note here that the systems in \cite{leleu_destabilization_2019}, \cite{goto_combinatorial_2019}, and \cite{molnar_continuous-time_2018} follow their own dynamics and are not related to the method of Lagrange multipliers. The system in \cite{leleu_destabilization_2019} will be discussed in Section~\ref{Ising solvers} in the main text while the work in \cite{goto_combinatorial_2019} will be discussed in Appendix D.

In effect, physical machines perform local steepest descent in the entropy generation rate. They can become stuck in local optima. At the very least, they perform a rapid search for local optima, thus reducing the search space for the global optimum. These machines are also adaptable toward searching in an expanded phase space, and other techniques for approaching a global optimum.

The paper is organized as follows. In Section~\ref{opti_phy}, we recognize that physics performs optimization through its variational and optimization principles. Then, we concentrate on the principle of minimum entropy generation or minimum power dissipation. In Section~\ref{coupled}, we give an overview of the minimum entropy generation optimization solvers in the literature and show how they incorporate constraints. Section~\ref{Lagrange tut} has a quick tutorial on the method of Lagrange Multipliers. Section~\ref{Ising solvers} studies five published solvers in detail and shows that they all follow some form of Lagrange multiplier dynamics. In Section~\ref{other}, we look at those published physics-based solvers which are less obviously connected to Lagrange multipliers. Section~\ref{reg} presents the applications of physics-based solvers to perform linear regression in statistics. Finally, in Section~\ref{conc}, we conclude and discuss the consequences of this ability to implement physics-based Lagrange multiplier optimization for areas such as machine learning.

%%%%%%%%%%%%%%%%%%%  Section 2 %%%%%%%%%%%%%%%%%%%%%%%%%%%%%%%%%%%%%%%%%%%%%%%%

\section{Optimization in Physics}
\label{opti_phy}
We survey the minimization principles of physics and the important optimization algorithms derived from them. These physical optimization machines are intended to converge to optima that are `agnostic to initial conditions'. By `agnostic to initial conditions', we mean systems that converge to the global optimum, or a good local optimum, irrespective of the initial point for the search. 

\subsection{Physics Principles and Algorithms}

\subsubsection{The principle of Least Action}
\begin{figure}[h]
\begin{center}
\includegraphics[scale=0.45]{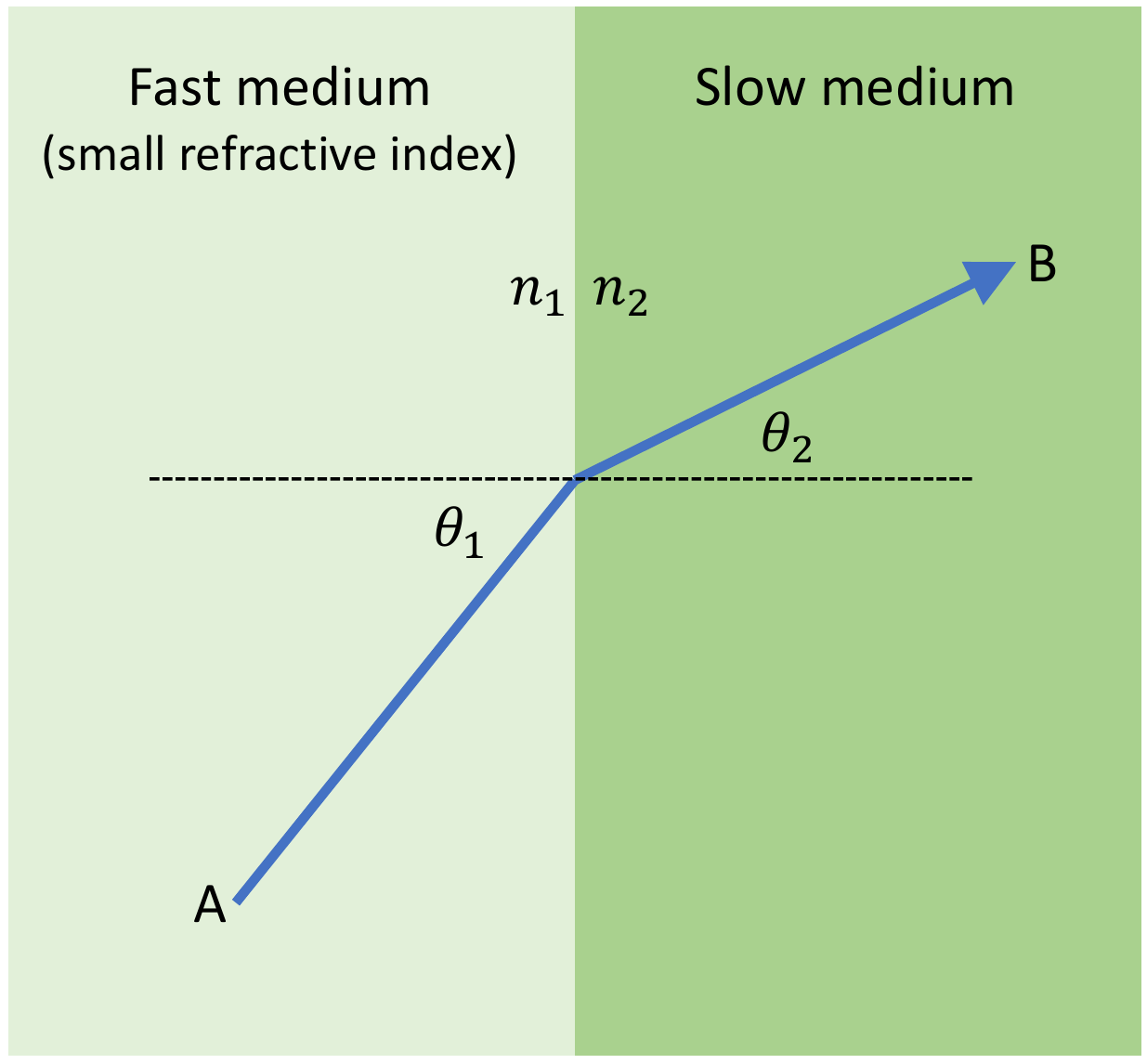}
\end{center}
\caption{\footnotesize \label{fig:LeastTime} The principle of Least Time, a subset of the principle of Least Action: The actual path that light takes to travel from point A to point B is the one that takes the least time to traverse. Recording the correct path entails a small energy cost consistent with the Landauer Limit.}
\end{figure}

The principle of Least Action is the most fundamental principle in physics. Newton's Laws of Mechanics, Maxwell's Equations of Electromagnetism, Schrödinger's equation in Quantum Mechanics, and Quantum Field Theory can all be interpreted as minimizing a quantity called action. For the special case of light propagation, this reduces to the principle of Least Time, as shown in Fig.~\ref{fig:LeastTime}.

A conservative system without friction or losses evolves according to the principle of Least Action. The fundamental equations of physics are reversible. A consequence of this reversibility is the Liouville Theorem which states that volumes in phase space are left unchanged as the system evolves.

Contrary-wise, in both a computer and an optimization solver, the goal is to have a specific solution with less uncertainty or a smaller zone in phase space than the initial state, an entropy cost first specified by Landauer and Bennett. Thus, some degree of irreversibility, or energy cost, is needed, specified by the number of digits in the answer in the Landauer/Bennett analysis. An algorithm has to be designed and programmed into the reversible system (either via software or via hardcoding of hardware) to effect the reduction in entropy needed to solve the optimization problem. Coming up with a fast algorithm for NP-hard problems is still an open problem in the field of reversible computing, which includes quantum computing.

This would require an energy cost, but not necessarily a requirement for continuous power dissipation. We look forward to computer science breakthroughs that would allow the principle of Least Action to address unsolved problems. An alternative approach to computing would involve physical systems that continuously dissipate power, aiding in the contraction of phase space toward a final solution. While exceeding the Landauer limit, such systems might have the advantage of speed and simplicity. This brings us to the principle of Least Power dissipation. 

\subsubsection{The principle of Least Power Dissipation}

\begin{figure}[h]
\begin{center}
\includegraphics[scale=0.35]{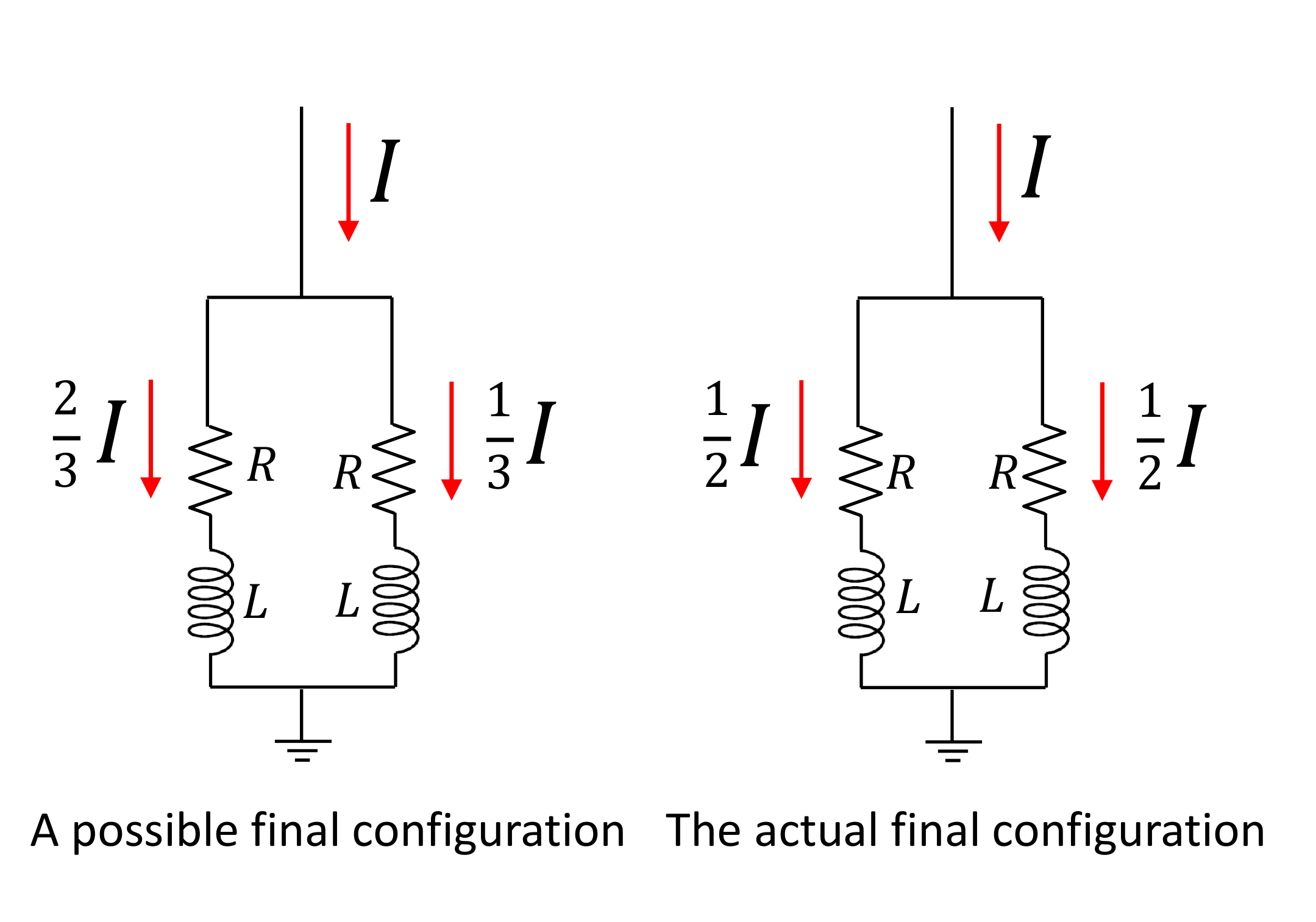}  
\end{center}
% Here is how to import EPS art
\caption{\footnotesize \label{fig:powerdiss} The principle of Least Power Dissipation: In a parallel connection, the current distributes itself in a manner that minimizes the power dissipation subject to the constraint of fixed input current $I$.}
\end{figure}

If we consider systems that continuously dissipate power, we are led to a second optimization principle in physics, the principle of Least Entropy Production or Least Power Dissipation. This principle states that any physical system will evolve into a steady-state configuration that minimizes the rate of entropy production given the constraints (such as fixed thermodynamic forces, voltage sources, or input power) that are imposed on the system. An early version of this statement is provided by Onsager in his celebrated papers on the reciprocal relations \cite{onsager_reciprocal_1931}. This was followed by further foundational work on this principle by Prigogine, \cite{prigogine_thermo_1947}, and de Groot, \cite{degroot_thermo_1951}. This principle is readily seen in action in electrical circuits and is illustrated in Fig.~\ref{fig:powerdiss}. We shall frequently use this principle, as formulated by Onsager, in the rest of the paper.

\subsubsection{Physical Annealing; Energy Minimization}
This technique is widely used in materials science and metallurgy and involves the slow cooling of a system starting from a high-temperature. As the cooling proceeds, the system tries to maintain thermodynamic equilibrium by reorganizing itself into the lowest energy minimum in its phase space. Energy fluctuations due to finite temperatures help the system escape from local optima. This procedure leads to global optima when the temperature reaches zero in theory but the temperature has to be lowered prohibitively slowly for this to happen. 
\begin{figure}[h]
\begin{center}
\includegraphics[scale=0.35]{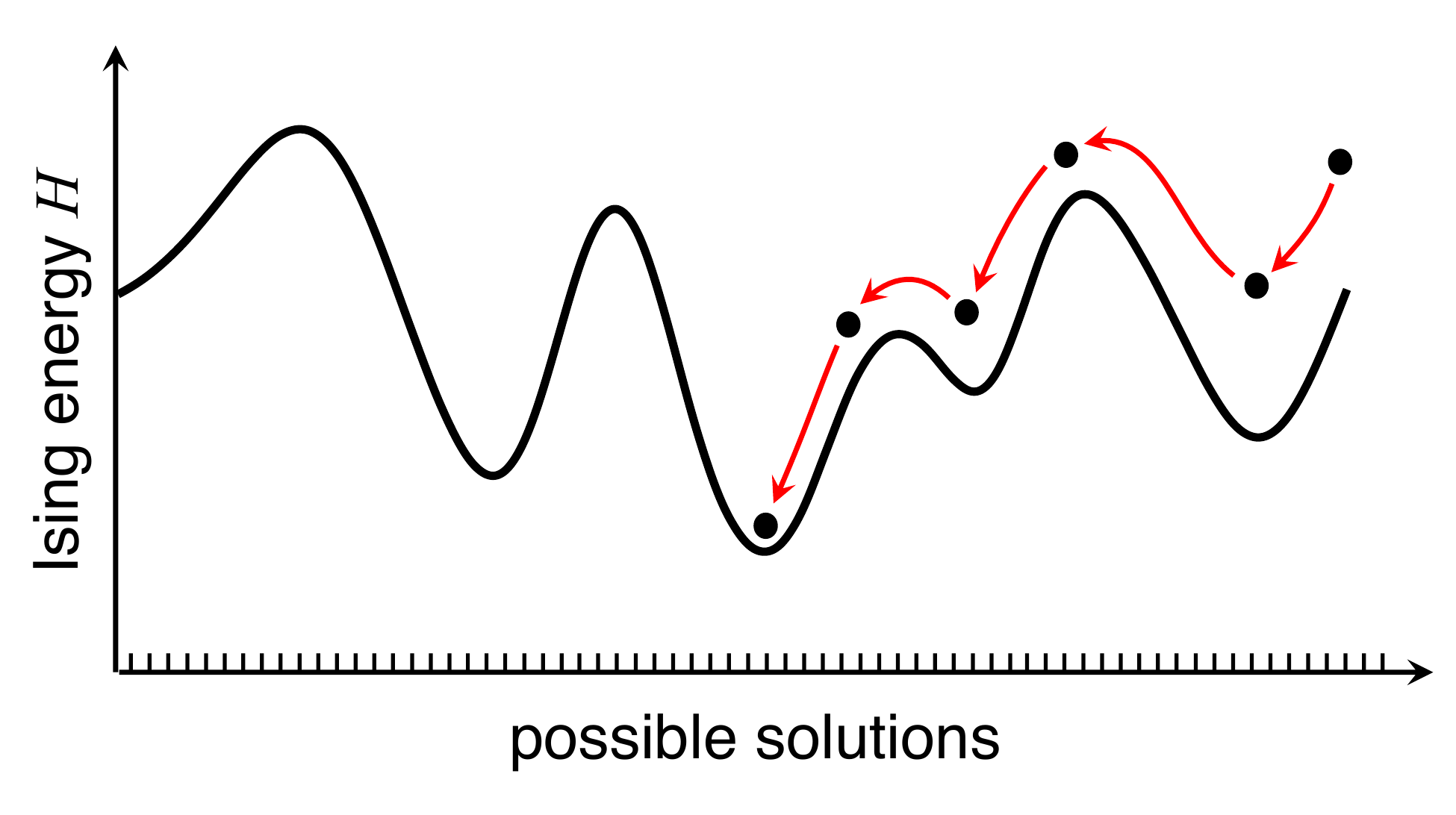}
\end{center}
% Here is how to import EPS art
\caption{
\footnotesize \label{fig:phyanneal} 
Physical annealing involves the slow cooling down of a system. The system performs gradient descent in configuration space with occasional jumps activated by finite temperature. If the cooling is done slowly enough, the system ends up in the ground state of configuration space.}
\end{figure}

\subsubsection{Adiabatic Method}
The Adiabatic Method involves the slow transformation of a system from initial conditions that are easily constructed to final conditions that capture the difficult problem at hand.

More specifically, to solve the Ising problem using this algorithm, one initializes the system of spins in the ground state of a simple Hamiltonian and then slowly varies the parameters of the Hamiltonian to end up with the final Ising Hamiltonian of interest. If the parameters are varied slowly enough, the system ends up in the ground state of the final Hamiltonian and the problem gets solved. In a quantum mechanical system, this is sometimes called `quantum annealing'. Several proposals and demonstrations, including the well-known D-Wave machine \cite{dickson_thermally_2013}, utilize this algorithm.

The slow rate of variation of the Hamiltonian parameters that is required for this method to work is determined by the minimum energy spacing between the instantaneous ground state and the instantaneous first excited state that occurs as we move from the initial Hamiltonian to the final Hamiltonian. The smaller the gap is for a particular variation schedule, the slower the rate at which we need to perform the variation to successfully solve the problem. It has been shown that the gap can get exponentially small in the worst case, implying that this algorithm can take exponential time in the worst case for \textbf{NP}-hard problems.

\begin{figure}[h]
\begin{center}
\includegraphics[scale=0.4]{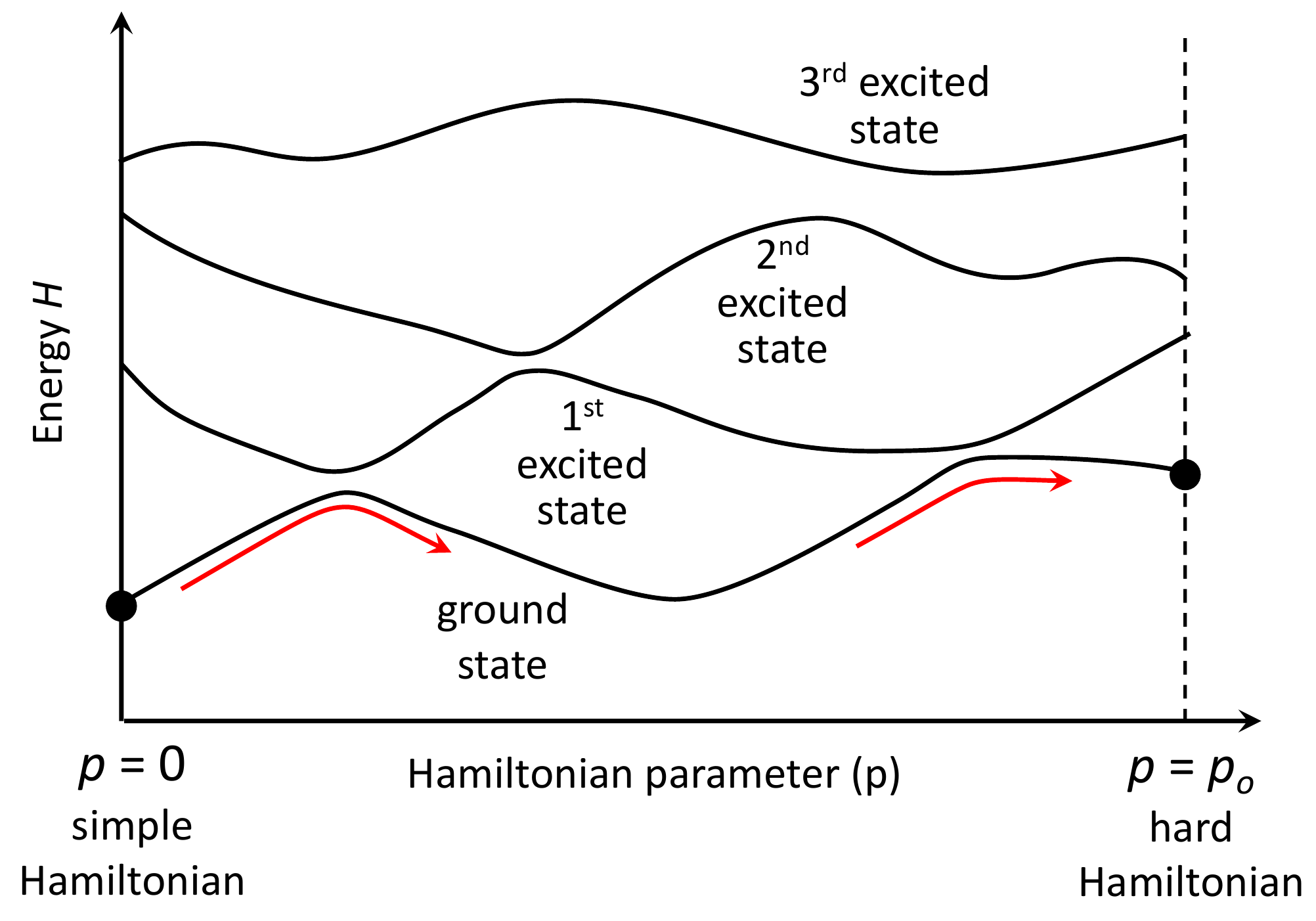}
\end{center}
% Here is how to import EPS art
\caption{\footnotesize \label{fig:adiabatic} A system initialized in the ground state of a simple Hamiltonian continues to stay in the ground state as long as the Hamiltonian is changed slowly enough.}
\end{figure}

\subsubsection{Minimum Entropy Generation in Multi-Oscillator Arrays}
Multi-Oscillator Arrays, subject to Parametric Gain were introduced in \cite{utsunomiya_mapping_2011} and \cite{haribara_computational_2016} for solving Ising problems. This can be regarded as a subset of the Principle of Minimum Entropy Generation, which is always subject to a non-zero input power constraint. In this case, gain acts as a boundary condition or constraint for the principle of minimum entropy generation, and the oscillator array must arrange itself to dissipate the least power subject to that constraint. In the context of a multi-coupled-oscillator arrays with gain, a certain oscillator mode will have the least loss. That mode will grow in amplitude most rapidly. This least loss mode can be individually selected by increasing the gain until it matches the intrinsic loss of that mode. Then further nonlinear evolution amongst all the modes will occur. If the oscillator array is bistable, as is the case for parametric gain which drives oscillation along the $\pm$real axis, this becomes the analog of magnetic bistability in an Ising problem. Then, we seek a solution for the lowest magnetic energy state with some oscillators locked at zero phase shift and others locked at $\pi$-phase shift. This mechanism will be the main point of Section~\ref{coupled}.

\section{Coupled Multi-Oscillator Array Ising Solvers}
\label{coupled}
The motivation for `Coupled Multi-Oscillator Array Ising Solvers' is best explained using concepts from laser physics. As a laser is slowly being turned on, spontaneous emission from the laser gain medium couples into the various cavity modes and begins to become amplified. The different optical modes in the cavity have different loss coefficients due to their differing spatial profiles. As the laser pump/gain increases, the cavity mode with the least loss grows faster than the other modes. Once the gain reaches the threshold gain, then further nonlinear evolution amongst all the modes will occur.

\begin{figure}[h]
\begin{center}
\includegraphics[scale=0.5]{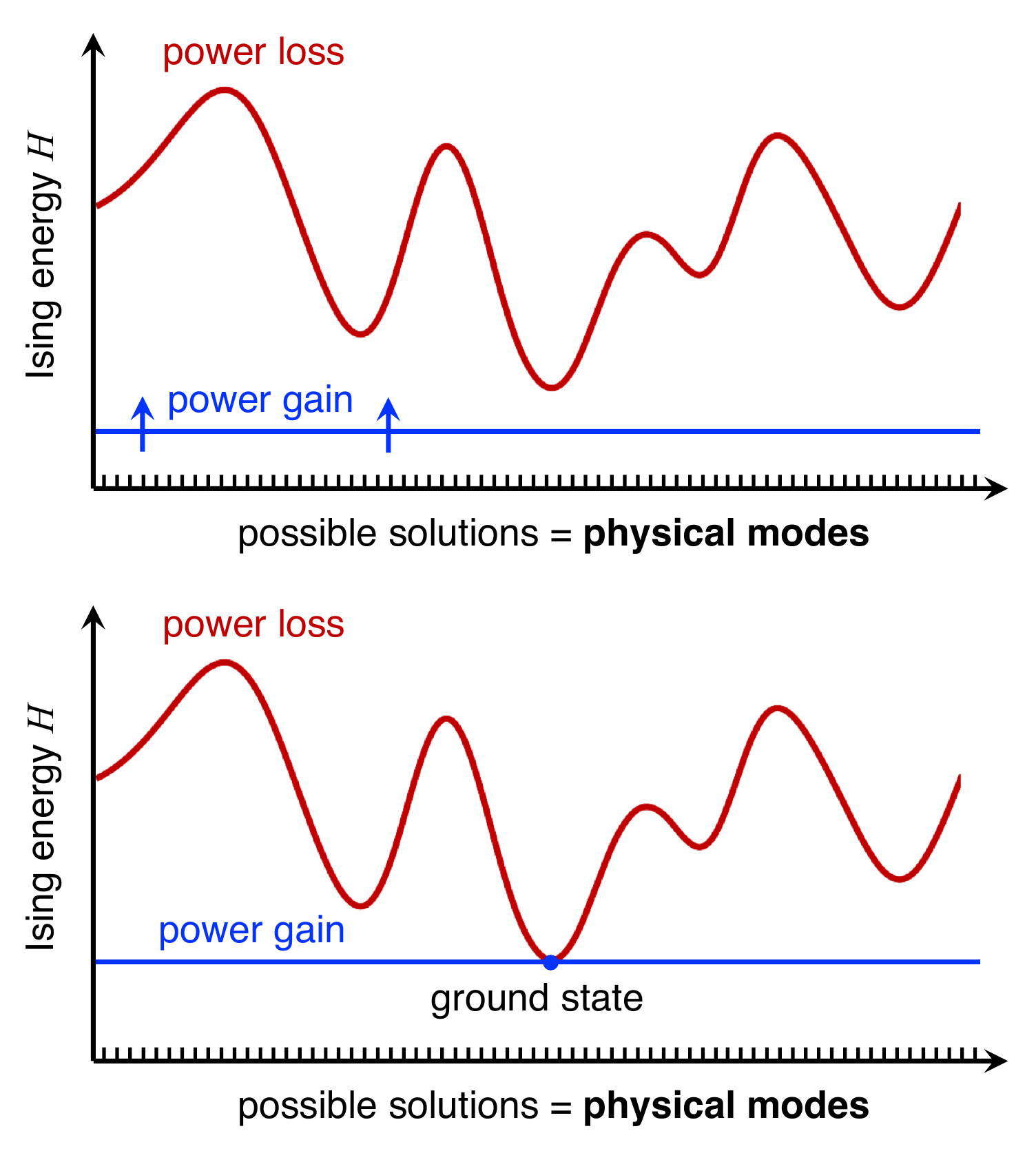}
\end{center}
% Here is how to import EPS art
\caption{\footnotesize \label{fig:motiv} A lossy multi-oscillator system is provided with gain: The x-axis is a list of all the available modes in the system whereas the y-axis plots the loss coefficient of each mode. Gain is provided to the system and is gradually increased. As in single-mode lasers, the lowest loss mode, illustrated by the blue dot, grows exponentially, saturating the gain. Above threshold we can expect further nonlinear evolution among the modes so as to minimize power dissipation.}
\end{figure}

The design of the Coupled Multi-Oscillator Array Ising machines tries to map the power losses of the optimization machine to the magnetic energies of various states
of the Ising problem. If the mapping is correct, the lowest power configuration will match the energetic ground state of the Ising problem. This is illustrated in Fig.~\ref{fig:motiv}. In effect, this is an example of a system evolving toward a state of minimum entropy generation, or minimum power dissipation, subject to the constraint of gain being present.

The archetypal solver in this class consists of a network of interconnected oscillators, driven by phase-dependent parametric gain. Parametric gain amplifies only the cosine quadrature and causes the electric field to lie along the $\pm$Real axis in the complex plane. The phase of the electric field (0 or $\pi$) can be used to represent $\pm$spin in the Ising problem. The resistive interconnections between the oscillators are designed to favor ferromagnetic or anti-ferromagnetic `spin-spin' interactions by the Principle of Minimum Entropy Generation, subject to the constraint of parametric (phase-dependent) gain as the power input. The parametric gain favors oscillation along the real axis of the complex plane, where the positive real axis would correspond to spin up, and the negative real axis would correspond to spin down.

The Voltage (or Current) input constraint is very important to the Principle of Minimum Entropy Generation. If there were no power input constraint, all the currents and voltages would be zero, and the minimum power dissipated would be zero. In the case of the Coupled Multi-Oscillator circuit, the power input is produced through a gain mechanism, or a gain module. The constraint could be the voltage input to the gain module. But if the gain were to be too small, it might not exceed the corresponding circuit losses, and the current and voltage would remain near zero. Thus, there is usually a threshold gain requirement, when applying the Principle of Minimum Entropy Generation to the Coupled Multi-Oscillator circuit.

If the gain is insufficient, the circuit will achieve minimum entropy generation at negligible currents and voltages. The intrinsic losses in the network would dominate and the circuit currents and voltages would be near zero. If the pump gain is then gradually ramped up the oscillatory mode requiring the least threshold gain begins oscillating. Upon reaching the threshold gain, a non-trivial current distribution built on the bistability of the Couple Multi-Oscillator circuit will emerge. As the gain exceeds the required threshold, there will be further nonlinear evolution among the modes so as minimize power dissipation. The final state “spin” configuration, dissipating the lowest power, (or entropy generation) emerges as the sought-for optimum.

Ideally, the gain evolution will be controlled by the Lagrange function to find the local minimum power dissipation configuration as will be discussed in Section~\ref{cambridge}. With Minimum Entropy Generation, as with most optimization schemes, it is difficult to guarantee a global optimum.

In optimization, each constraint contributes a Lagrange multiplier. We will show that the gains of the oscillators are the Lagrange multipliers of the constrained system. In the next section, we provide a brief tutorial on Lagrange Multiplier optimization.
%%%%%%%%%%%%%%%%%%%%%%%%%%%%%%%%%%%%%%%%%%%%%%%%%%%%%%%%%%%%%%%%%%%%%%%%%%%%%%%%%

\section{Lagrange Multiplier Optimization Tutorial}
\label{Lagrange tut}
\begin{figure}[h]
\begin{center}
\includegraphics[scale=0.43]{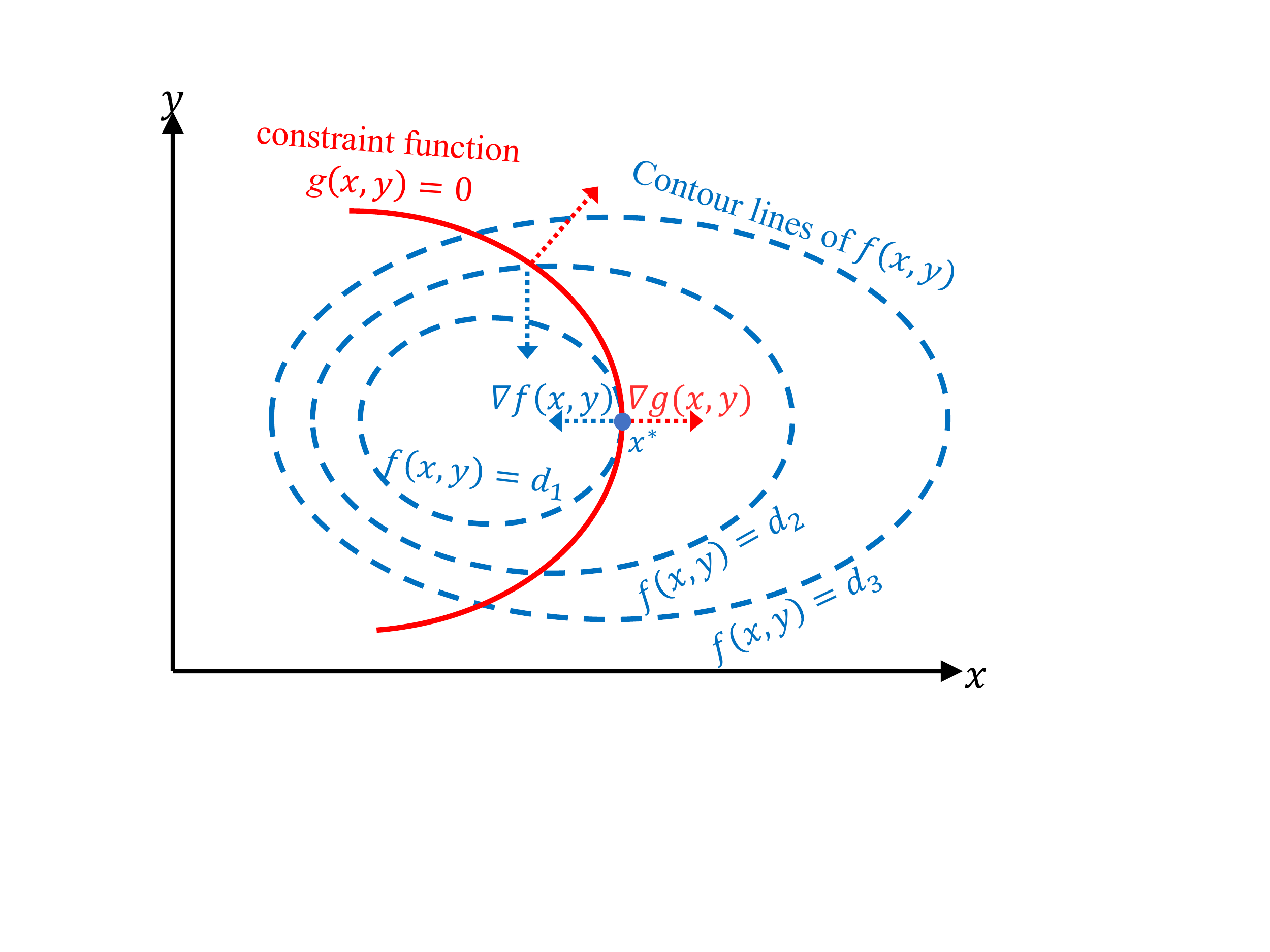} 
\end{center}
% Here is how to import EPS art
\caption{\footnotesize \label{fig:lag} Maximization of function $f(x,y)$ subject to the constraint $g(x,y)=0$. At the constrained local optimum, the gradients of $f$ and $g$, namely $\nabla f(x,y)$ and $\nabla g(x,y)$, are parallel.}
\end{figure}

The method of Lagrange multipliers is a very well-known procedure for solving constrained optimization problems. In constrained optimization, the optimal point $\bm{x^*}\equiv(x,y)$ in multi-dimensional solution space locally optimizes the merit function $f(\bm{x})$ subject to the constraint $g(\bm{x})=0$. The optimal point has the property that the slope of the merit function is zero as infinitesimal steps are taken away from $\bm{x^*}$, as taught in calculus. But these deviations are restricted to the constraint curve, as shown in Fig.~\ref{fig:lag}. The iso-contours of the function $f(\bm{x})$ increase until they are limited by, and just touch, the constraint curve $g(\bm{x})=0$ at the point $\bm{x^*}$. 

At the point of touching, $\bm{x^*}$, the gradient of $f$ and the gradient of $g$ are parallel to each other. This can be stated formally as:
\begin{equation*}
    \bm{\nabla} f(\bm{x^*})=\lambda^*\bm{\nabla} g(\bm{x^*}).
\end{equation*}
The proportionality constant $\lambda^*$ is called the Lagrange multiplier corresponding to the constraint $g(\bm{x})=0$. 

Although Fig.~\ref{fig:lag} shows only a 2-dimensional optimization space, in general let us take the optimization space $x$ to be $n$-dimensional. When we have multiple constraints $g_1,\dots,g_p$, we correspondingly expand the $n$-dimensional Lagrange Multiplier requirement:
\begin{equation*}
    \bm{\nabla} f(\bm{x^*})=\sum_{i=1}^p\lambda_i^*\bm{\nabla} g_i(\bm{x^*}),
\end{equation*}
where the gradient vector $\bm{\nabla}$ represents $n$-equations, accompanied by the $p$ constraint equations $g_i(\bm{x})=0$, resulting in $n+p$ equations. These equations solve for the $n$ components in the vector $\bm{x^*}$, and the $p$ unknown Lagrange Multipliers $\lambda_i^*$. That would be $n+p$ equations for $n+p$ unknowns.

Motivated by the above condition, let us introduce a Lagrange function $L(\bm{x},\bm{\lambda})$ defined as follows:
\begin{equation*}
    L(\bm{x}, \bm{\lambda}) = f(\bm{x}) + \sum_{i=1}^p\lambda_ig_i(\bm{x}),
\end{equation*}
which can be optimized by gradient descent or other methods to solve for $\bm{x^*}$ and $\bm{\lambda^*}$. The full theory of Lagrange multipliers, and the popular `Augmented Lagrangian Method of Multipliers' algorithm used to solve for locally optimal $(\bm{x^*}, \bm{\lambda^*})$, are discussed in great detail in \cite{boyd_convex_2004} and \cite{bertsekas_nonlinear_1999}. A gist of the main points is presented in Appendices A, B, C.

For the specific case of the Ising problem, the objective function is given by $f(\bm{\mu})=\nobreak\sum_{i,j}J_{ij}\bm{\mu_i}\cdot\bm{\mu_j}$, where $f(\bm{\mu})$ is the magnetic Ising Energy and $\bm{\mu_i}$ is the $i$-th magnetic moment vector. For the optimization method represented in this paper, we need a circuit or other physical system whose power dissipation is also $f(\bm{x})=\nobreak\sum_{i,j}J_{ij}x_ix_j$, but now $f(\bm{x})$ is power dissipation, not energy, $x_i$ is a variable that represents voltage, or current or electric field, and the $J_{ij}$ are not magnetic energy, but rather resistive coupling elements. The correspondence is between magnetic spins quantized along the z-axis, $\mu_{zi}=\pm 1$, and the circuit variable $x_i=\pm 1$.

While `energy' as opposed to `power dissipation' are represented by different units, nonetheless we need to establish a correspondence between them. For every optimization problem, there is a challenge of finding a physical system whose power dissipation function represents the desired equivalent optimization function.

If the Ising problem has $n$ spins, there are also $p=n$ constraints, one for each of the spins. A sufficient constraint is: $g_i(\bm{x})=1-x_i^2$. More complicated nonlinear constraints can be envisioned, but $(1-x_i^2)$ could represent the first two terms in a more complicated constraint Taylor expansion.

Therefore, a sufficient Lagrange function for the Ising problem, with digital constraints, is given by:
\begin{equation*}
    L(\bm{x},\bm{\lambda})=\sum_{i=1}^n\sum_{j=1}^nJ_{ij}x_ix_j+\sum_{i=1}^n\lambda_i\lp1-x_i^2\rp,
\end{equation*}
where $\lambda_i$ is the Lagrange Multiplier associated with the corresponding constraint. We shall see in the next section that most analog algorithms that have been proposed for the Ising problem in the literature, actually tend to optimize some version of the above Lagrange function.
%%%%%%%%%%%%%%%%%%%%%%%%%%%%%%%%%%%%%%%%%%%%%%%%%%%%%%%%%%%%%%%%%%%%%%%%%%%%%%%%%%%

%%%%%%%%%%%%%%%%%% Section 3 %%%%%%%%%%%%%%%%%%%%%%%%%%%%%%%%%%%%%%%%%%%%%%%%%%%%

\section{The Physical Ising Solvers}
\label{Ising solvers}
In this section, we discuss each physical procedure proposed in the literature and show how each physical scheme implements the method of Lagrange multipliers. They all obtain good performance on the Gset benchmark problem set \cite{gset}, and many of them demonstrate that they perform better than the heuristic algorithm, Breakout Local Search \cite{benlic_breakout_2013}. The main result of our work is the realization that the pump gains used in all the physical methods are in fact Lagrange multipliers.

The available physical solvers in the literature are as follows:
\begin{enumerate}
    \item Optical Parametric Oscillators
    \item Coupled Radio Oscillators on the Real Axis
    \item Coupled laser cavities using multicore fibers
    \item Coupled Radio Oscillators on the Unit Circle
    \item Coupled polariton condensates
\end{enumerate}
Then there are a number of schemes that also rely upon a variant of minimum entropy production or power dissipation:
\begin{enumerate}
    \item[6.] Iterative Analog Matrix Multipliers
    \item[7.] Leleu Mathematical Ising Solver
\end{enumerate}
In Appendix D, there is scheme that appears unconnected with minimum entropy production rate:
\begin{enumerate}
    \item[8.] Adiabatic coupled radio oscillators (Toshiba)
\end{enumerate}

We shall see that methods 1, 2, 4, in the literature use only one gain for all the oscillators which is equivalent to imposing only one constraint. The other methods, 3, 5, 6, use different gains for each spin and correctly capture the full $n$ constraints of the Ising problem.

\subsection{Optical Parametric Oscillators}
\label{yamamoto}
\subsubsection{Overview}
An early optical machine for solving the Ising problem was presented by Yamamoto et al. \cite{utsunomiya_mapping_2011} and \cite{mcmahon_fully_2016}. Their system consists of several pulses of light circulating in an optical fiber loop, with the phase of each light pulse representing an Ising spin.
In parametric oscillators, gain occurs at half the pump frequency. If the gain overcomes the intrinsic losses of the fiber, the optical pulse builds up. Parametric amplification provides phase dependent gain. It restricts the oscillatory phase to the Real Axis of the complex plane. This leads to bistability along the positive or negative real axis, allowing the optical pulses to mimic the bistability of magnets. 

In the Ising problem, there is magnetic coupling between spins. The corresponding coupling between optical pulses is achieved by controlled optical interactions between the pulses. In Yamamoto’s approach, one pulse $i$ is first plucked out by an optical gate, amplitude modulated by the proper connection weight specified in the $J_{ij}$ Ising Hamiltonian, and then reinjected and superposed onto the other optical pulse $j$, producing constructive or destructive interference, representing ferromagnetic or anti-ferromagnetic coupling.

By providing saturation to the pulse amplitudes, the optical pulses will finally settle down, each to one of the two bistable states. We will find that the pulse amplitude configuration evolves exactly according to the Principle of Minimum Entropy Generation. If the magnetic dipole solutions in the Ising problem are constrained to $\pm 1$ then each constraint is associated with a Lagrange Multiplier. We find that each Lagrange Multiplier turns out to be equal to the gain or loss associated with the corresponding oscillator.

\subsubsection{Lagrange multipliers as gain coefficients}
As an example, Yamamoto et al. \cite{haribara_computational_2016} analyze their parametric oscillator system using coupled wave equations for the slowly varying amplitudes of the circulating optical modes. We now show that the coupled wave equation approach reduces to an extremum of their system `Entropy Generation' or `power dissipation'. The coupled-wave equation for parametric gain of the slowly varying amplitude $c_i$ of the in-phase cosine component of the $i$-th optical pulse (representing magnetic spin in an Ising system) is as follows:
\begin{equation}
    \frac{dc_i}{dt}=(-\alpha_i+\gamma_i)c_i - \sum_j J_{ij}c_j, \label{Eq1}
\end{equation}
where the weights, $J_{ij}$, are the magnetic ferromagnetic or anti-ferromagnetic cross-couplings, and $\gamma_i$ represents the phase dependent parametric gain given by the pump to the $i$-th circulating pulse, whereas $\alpha_i$ is the corresponding dissipative loss. The amplitudes $c_i$ represent the optical cosine wave electric field amplitudes. In the case of circuits, the voltage amplitudes can be expressed as $V_i = c_i+\text{j}s_i$ where $c_i$ and $s_i$ represent the cosine and sine quadrature components of voltage, and j is the unit imaginary. For clarity of discussion, we dropped the cubic terms in \eqref{Eq1} that Yamamoto et al originally had. A discussion of these terms in given in Appendix C.

Owing to the nature of parametric amplification, the quadrature sine wave components $s_i$ of the electric field amplitude die out rapidly. The rate of entropy generation, or net power dissipation $h$, including the negative dissipation associated with gain can be written:
\begin{equation}
    h(c_1,\dots,c_n)=\sum_{i,j}J_{ij}c_ic_j+\sum_i\alpha_ic_i^2-\sum_i\gamma_ic_i^2, \label{2}
\end{equation}
If we minimize the entropy generation $h(\bm{c})$ without invoking any constraints, that is, with $\gamma_i=0$, the amplitudes $c_i$ simply go to zero, which generates the minimum entropy.

If the gain $\gamma_i$ is large enough, some of the amplitudes might go to infinity. To avoid this, we may employ the $n$ constraint functions $g_i(c_i)=\lp1-c_i^2\rp=0$, which enforce a digital $c_i=\pm 1$ outcome. (Actually, a constraint of the form $g_i(c_i)=\lp1-c_i^2\rp=0$ is quite general in the sense that $\lp1-c_i^2\rp$ can represent the first two terms of the Taylor Series of an arbitrary  constraint.) Adding the constraint function to the entropy generation, yields the Lagrange function including the constraint functions times the respective Lagrange Multipliers:
\begin{equation}
    L(\bm{c},\bm{\gamma})=\sum_{i,j}J_{ij}c_ic_j+\sum_i\alpha_ic_i^2-\sum_i\gamma_i\lp c_i^2-1\rp, \label{3}
\end{equation}
Comparing the unconstrained \eqref{2} to the constrained \eqref{3}, they only differ in the final (-1) term which effectively constrains the amplitudes and prevents them from diverging to $\infty$. Expression \eqref{3} is the Lagrange function given at the end of Section~\ref{Lagrange tut}. Surprisingly, the gains $\gamma_i$ emerge to play the role of Lagrange Multipliers. This means that each mode, represented by the subscripts in $c_i$, must adjust to a particular gain $\gamma_i$ which minimizes the overall entropy generation, and the respective gains $\gamma_i$ represent the Lagrange Multipliers. Minimization of the Lagrange function \eqref{3} provides the final steady state of the system dynamics.

If the circuit or optical system is designed to dissipate power, or equivalently generate entropy, in a mathematical form that matches the magnetic energy in the Ising problem, then the dissipative system will seek out a corresponding local optimum configuration of the magnetic Ising energy.

Such a physical system, constrained to $c_i=\pm 1$, is digital in the same sense as a flip-flop circuit, but unlike the von Neumann computer, the inputs are resistor weights for power dissipation. Nonetheless a physical system can evolve in a direct manner, without the need for shuttling information back and forth as in a von Neumann computer, providing faster answers. Without the communications overhead but with the higher operation speed, the energy dissipated to arrive at the final answer will be less, in spite of the circuit being required to generate entropy during its evolution toward the final state.

To achieve minimum entropy production, the amplitudes $c_i$, and the Lagrange Multipliers $\gamma_i$, must all be simultaneously optimized. While a circuit will evolve toward optimal amplitudes $c_i$, the gains $\gamma_i$ must arise from a separate active circuit. Ideally, the active circuit which controls the Lagrange Multiplier gains $\gamma_i$, would have its entropy production included with the main circuit. A more common method is to provide gain that follows a heuristic rule using an external feedback circuit. For example, Yamamoto et al. follow the heuristic rule $\gamma_i =a+bt$. It is not yet clear whether the heuristic-based approach toward gain evolution will be equally effective as simply lumping together all main circuit and feedback components and simply minimizing the total power dissipation.

We conclude this subsection by noting that the Lagrangian, Eq \eqref{3}, corresponds to Lagrange multiplier optimization using the following merit function and constraints:
\begin{gather*}
    f(\bm{c})=\sum_{i,j}J_{ij}c_ic_j+\sum_i\alpha_ic_i^2,\\
    g_i(c_i)=\lp1-c_i^2\rp=0.
\end{gather*}
\subsubsection{Conclusions from Optical Parametric Oscillators}
\begin{enumerate}
    \item Physical systems minimize the entropy production rate, or power dissipation, subject to input constraints of voltage, amplitude, gain, etc.
    \item These systems actually perform Lagrange Multiplier optimization.
    \item Indeed, it is the gain $\gamma_i$ in each oscillator $i$ that plays the role of the corresponding Lagrange Multiplier.
    \item If the Lagrange function is split in such a way that only the Ising function, $\sum_{i,j}J_{ij}c_ic_j$, is treated as the merit function $f(\bm{c})$, then the Lagrange multipliers corresponding to the constraints $g_i$ are the \emph{net} gains, $\gamma_i-\alpha_i$.    
    \item Under the digital constraint, amplitudes $c_i=\pm 1$, entropy generation minimization schemes are actually binary, similar to a flip-flop.
\end{enumerate}

\subsection{Coupled Radio Oscillators on the real axis}
\label{patrick}
\subsubsection{Overview}
\begin{figure}[h]
\begin{center}
\includegraphics[scale=0.4]{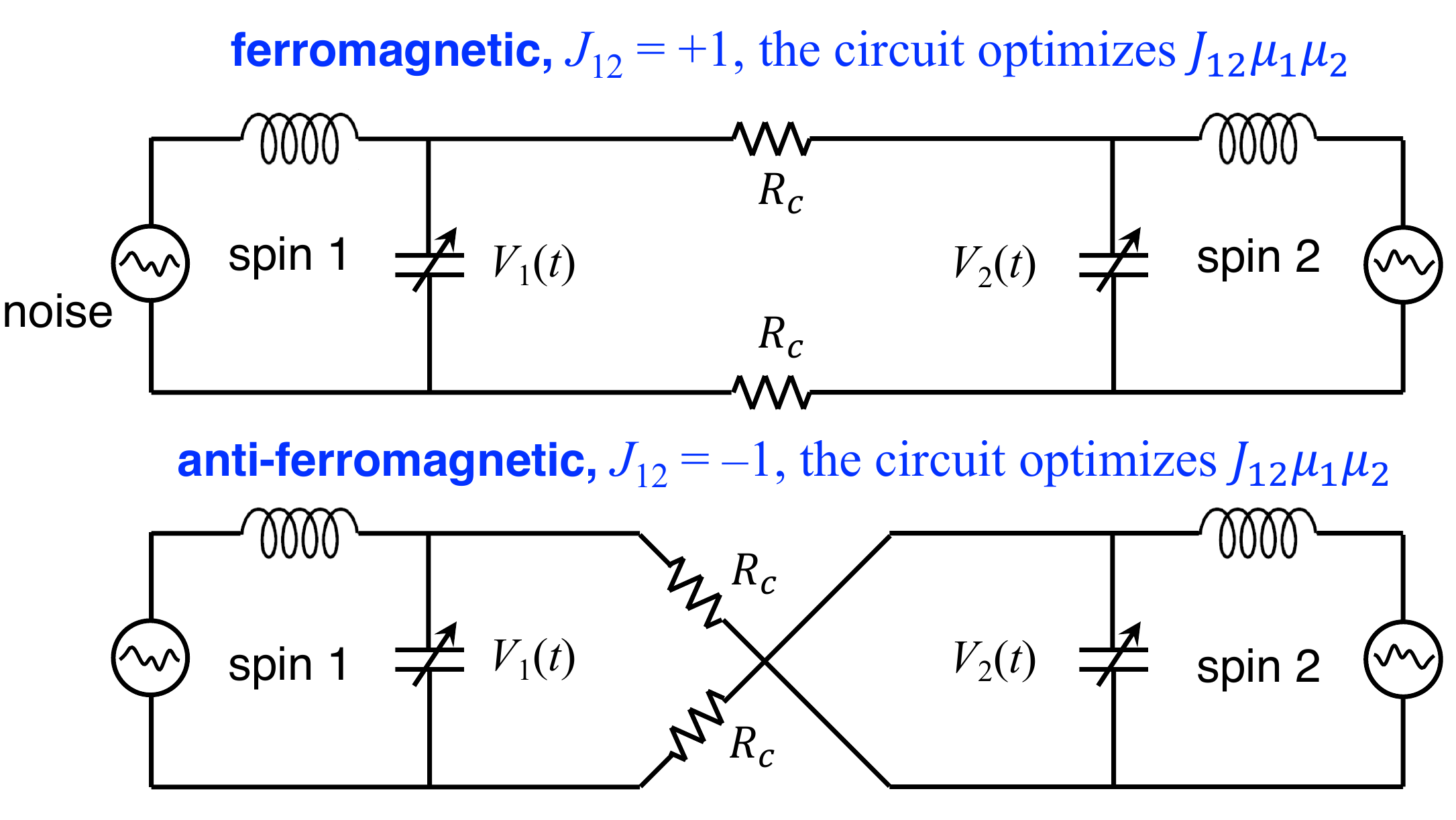}
\end{center}
% Here is how to import EPS art
\caption{\footnotesize \label{fig:patrick} Coupled LC oscillator circuit for two coupled magnets. The oscillation of the LC oscillators represents the magnetic moments while the parallel or antiparallel cross-connections represent ferromagnetic or anti-ferromagnetic coupling, respectively. The nonlinear capacitors are pumped by $V(2\omega_0)$ at frequency $2\omega_0$ providing parametric gain at $\omega_0$.}
\end{figure}

A coupled LC oscillator system with parametric amplification was analyzed in the circuit simulator, SPICE, by Xiao et al., \cite{xiao_optoelectronics_2019}. This is analogous to the optical Yamamoto system but this system consists of a network of radio frequency LC oscillators coupled to one another through resistive connections. The LC oscillators contain linear inductors but nonlinear capacitors which provide the parametric gain. The parallel or cross-connect resistive connections between the oscillators are designed to implement the ferromagnetic or anti-ferromagnetic couplings $J_{ij}$ between magnetic dipole moments $\mu_i$ as shown in Fig.~\ref{fig:patrick}. The corresponding phase of the voltage amplitude $V_i$, 0 or $\pi$ determines the sign of magnetic dipole moment $\mu_i$.

The nonlinear capacitors are pumped by voltage $V(2\omega_0)$ at frequency $2\omega_0$, where the LC oscillator natural frequency is $\omega_0$. Second harmonic pumping leads to parametric amplification in the oscillators. As in the optical case, parametric amplification plays the dual role of generating/sustaining voltage oscillations as well as imposing phase bistability to the ac voltages in the oscillators. The bistability refers to gain along the $\pm$Real-Axis defined by time synchronization with the $2\omega_0$-pump.

The $2\omega_0$-pump induces gain $\gamma_i$ in the Real-Axis quadrature. As in the case of optical parametric amplifier machines, ideally, an active circuit would control the Lagrange Multiplier gains $\gamma_i$, and the gain control circuit would have its entropy production included with the main circuit. A more common approach is to provide gain that follows a heuristic rule.

As in the optical parametric amplifier case, a mechanism is needed to prevent the parametric gain from producing infinite amplitude signals. Zener diodes can be inserted into the circuit to restrict the amplitudes to finite saturation values or to digitally defined values. With the diodes in place, the circuit settles into a voltage phase configuration, 0 or $\pi$, that minimizes net power dissipation for a given pump gain.

\subsubsection{Lagrange function and Lagrange multipliers}
The amplitudes can be defined by the voltages across the LC oscillator capacitors and derived from Kirchoff's voltage and current equations as in Xiao, \cite{xiao_optoelectronics_2019}. Performing the slowly-varying amplitude approximation on the cosine component of these voltages, $c_i$, Xiao obtains the following equation of motion:
\begin{equation}
    \frac{dc_i}{dt}=\frac{1}{4R_cC_0}\sum_j{J_{ij}c_j}-\alpha_ic_i+\gamma_ic_i, \label{4}
\end{equation}
where the $c_i$ are the peak voltage amplitudes in units of a reference voltage, $R_c$ is the resistance of the coupling resistors, the cross-couplings $J_{ij}$ are assigned binary values $J_{ij}=\pm1$, $C_0$ is the linear part of the capacitance in each oscillator, $n$ is the number of oscillators, $\omega_0$ is the natural frequency of all the oscillators, and the parametric gain constant $\gamma=\omega_0|\Delta C|/4C_0$ where $|\Delta C|$ is the capacitance modulation at the second harmonic. In the decay constant $\alpha=(n-1)/(4R_cC_0)$, there are $(n-1)$ resistors $R_c$ in parallel, since it is assumed that each oscillator can give up energy to all the other $(n-1)$ oscillators, with the coupling resistors acting in parallel. In this simplified model all decay constants $\alpha$ are taken as equal, and moreover each oscillator experiences exactly the same parametric gain $\gamma$; conditions that can be relaxed if needed.

The number 4 is present in the first denominator, since for two coupled LC circuits as shown in Fig.~\ref{fig:patrick}, the decay is controlled by the RC time of the capacitors in parallel and the $R_c$ resistors in series. Likewise, the number 4 appears for loss and parametric gain, but for different reasons in each case.

We note that equation \eqref{4} above performs gradient descent on the net power dissipation function:
\begin{equation}
    h(\bm{c},\gamma)=-\frac{1}{4R_c}\sum_{i,j}J_{ij}c_ic_j+\sum_i \alpha C_0c_i^2- \sum_i \gamma C_0c_i^2, \label{5}
\end{equation}
which is very similar to Section~\ref{yamamoto}. On the right-hand side, the reason for the 4 in the first denominator is to compensate for double counting in $i\ne j$. The first two terms on the right-hand side together represent the dissipative losses in the coupling resistors while the third term is the negative of the gain provided to the system of oscillators.

Next, we obtain the following Lagrange function through the same replacement of $\lp-c_i^2\rp$ with $\lp1-c_i^2\rp$ that we performed in Section~\ref{yamamoto}:
\begin{equation}
    L(\bm{c},\gamma)=-\frac{1}{4R_c}\sum_{i,j}J_{ij}c_ic_j+\sum_i \alpha C_0c_i^2- \sum_i \gamma C_0\lp c_i^2-1\rp, \label{6}
\end{equation}
The above Lagrangian corresponds to Lagrange multiplier optimization using the following merit function and constraints:
\begin{gather*}
    f(\bm{c})=-\frac{1}{4R_c}\sum_{i,j}J_{ij}c_ic_j+\sum_i \alpha C_0c_i^2,\\
    g(\bm{c})=\sum_i\lp1-c_i^2\rp=0.
\end{gather*}
Again, we see that the gain coefficient $\gamma$ is the Lagrange multiplier of the constraint $g=0$.

\subsubsection{Iterative Optimization as a Series of Time Steps}
Although the extremum of Eq \eqref{6} represents the final evolved state of a physical system representing an optimization outcome, it would be interesting to examine the time evolution toward the optimal state. The optimization occurs by iterative steps, where each iteration can be regarded to take place in a time interval $\Delta t$. At each successive iteration, the voltage amplitude $c_i$ takes a step whose magnitude is proportional to the gradient of the Lagrange function:
\begin{equation}
    c_i(t+\Delta t)=c_i(t)-\kappa\Delta t\frac{\partial}{\partial c_i}L(\bm{c},\gamma), \label{7}
\end{equation}
where the minus sign on the right hand side drives the system toward minimum power dissipation. As the Lagrange function comes closer to its minimum, the gradient $\frac{\partial}{\partial c_i}L(\bm{c},\gamma)$ diminishes and the amplitude steps become smaller and smaller. The adjustable proportionality constant $\kappa$, controls the size of each iterative step; it also calibrates the dimensional units between power dissipation and voltage amplitude. (Since $c_i$ is voltage amplitude, $\kappa$ has units of reciprocal capacitance.) Converting Eq \eqref{7} to continuous time,
\begin{gather}
    \frac{dc_i}{dt}=-\kappa\frac{\partial}{\partial c_i}L(\bm{c},\gamma), \label{8}\\
    \frac{dc_i}{dt}=-\kappa\lp\frac{\partial}{\partial c_i}f(c_i)+\sum_j\gamma_j\frac{\partial}{\partial c_i}g_j(c_j)\rp,
\end{gather}
where the $\gamma_j$ play the role of Lagrange multipliers, and the $g_j=0$ are the constraints. Taking $L(\bm{c},\gamma)$ from Eq \eqref{6}, the gradient in the voltage amplitudes becomes
\begin{equation*}
    \frac{\partial}{\partial c_i}L(\bm{c},\gamma)=-\frac{1}{2R_c}\sum_j{J_{ij}c_j} +2\alpha C_0c_i-2\gamma C_0c_i.
\end{equation*}
Substituting into Eq \eqref{8}, the time derivative of the voltage amplitudes becomes
\begin{equation}
    \frac{dc_i}{dt}=2\kappa\lp\frac{1}{4R_c}\sum_{j}{J_{ij}c_j} -\alpha C_0c_i+\gamma C_0c_i\rp. \label{10}
\end{equation}
The constant $\kappa$ can be absorbed into the units of time to reproduce the dynamical equation \eqref{4}, the slowly varying amplitude approximation for the coupled radio oscillators. Thus, it is interesting that the slowly-varying time dynamics can be reproduced from iterative optimization steps on the Lagrange function.

\subsubsection{Conclusions from Coupled Oscillators on the Real Axis}
\begin{enumerate}
    \item As in other cases, the coupled LC oscillator system \cite{xiao_optoelectronics_2019} minimizes the entropy production rate, or power dissipation, incorporating the power input from the pump gain.
    \item The coupled LC oscillator system actually performs Lagrange Multiplier optimization whose Merit Function is the Lagrange Function.
    \item The gain $\gamma_i$ in each oscillator $i$ plays the role of the corresponding Lagrange Multiplier.
    \item Under the amplitude constraint on the voltage cosine component, $c_i=\pm 1$, the entropy generation minimization scheme is actually binary, similar to a flip-flop.
    \item Successive iterations toward a power dissipation minimum, employing gradient descent in the amplitudes of the cosine components of the voltages $c_i$ , surprisingly, yield the same time-dependent Slowly Varying amplitude differential equation \eqref{4}.
\end{enumerate}

\subsection{Coupled laser cavities using multicore fibers}
\label{arizona}
\subsubsection{Overview}
The Ising solver designed by Babaeian et al., \cite{babaeian_single_2019}, makes use of coupled laser modes in a multicore optical fiber. Polarized light in each core of the optical fiber corresponds to each magnetic moment in the Ising problem. This means that the number of cores needs to be equal to the number of magnets in the given Ising instance. The right-hand circular polarization and left-hand circular polarization of the laser light in each core represent the two polarities (up and down) of the corresponding magnet. The mutual coherence of the various cores is maintained by injecting seed light from a master laser.

The coupling between the fiber cores is achieved through amplitude mixing of the laser modes by Spatial Light Modulators at one end of the multicore fiber, \cite{babaeian_single_2019}. These Spatial Light Modulators couple light amplitude from the $i$-th core to the $j$-th core according to the prescribed connection weight $J_{ij}$.

\subsubsection{Equations and comparison with Lagrange multipliers}

As in prior physical examples the electric field amplitudes can be expressed in the slowly-varying polarization modes of the $i$-th core, $E_{iL}$ and $E_{iR}$, where the two electric field amplitudes are in-phase temporally, are positive real, but have different polarization. They are,
\begin{align*}
    \frac{d}{dt}E_{iL}=&-\alpha_iE_{iL}+\gamma_iE_{iL}+\frac{1}{2}\sum_j{J_{ij}\lp E_{jR}-E_{jL}\rp},\\
    \frac{d}{dt}E_{iR}=&-\alpha_iE_{iR}+\gamma_iE_{iR}-\frac{1}{2}\sum_j{J_{ij}\lp E_{jR}-E_{jL}\rp},
\end{align*}
where $\alpha_i$ is the decay rate in the optical core in 1/sec units, and $\gamma_i$ is the gain supplied to the $i$-th core. The first term on the right in both the equations represents optical fiber losses while the second term represents the
gain provided. The third term represents the coupling between the $j$-th and $i$-th cores that is provided by the Spatial Light Modulators. We next define the degree of polarization as $\mu_i\equiv E_{iL}-E_{iR}$. (This differs from the usual definition of degree of polarization of an optical beam which contains intensity rather than electric field amplitude.) Subtracting the two equations above, we obtain the following evolution equation for $\mu_i$:
\begin{equation*}
    \frac{d}{dt}\mu_i=-\alpha_i\mu_i+\gamma_i\mu_i+\sum_j{J_{ij}\mu_j}.
\end{equation*}
The power dissipation, or entropy generation, is proportional to two orthogonal components of electric field, each squared, $|E_{iL}|^2+|E_{iR}|^2$. But this can also be written $|E_{iL}-E_{iR}|^2+|E_{iL}+E_{iR}|^2= |\mu_i|^2+|E_{iL}+E_{iR}|^2$. But $|E_{iL}+E_{iR}|^2$ can be regarded as relatively constant as energy switches back and forth between right and left circular polarization. Then the changes in power dissipation, or entropy generation, $h(\bm{\mu})$ would be most influenced by quadratic terms in $\bm{\mu}$:
\begin{equation*}
    h(\bm{\mu},\bm{\gamma})=\sum_i\alpha_i\mu_i^2+\sum_{i,j}J_{ij}\mu_i\mu_j-\sum_i\gamma_i\mu_i^2.
\end{equation*}
As before, we add the $n$ digital constraints $g_i(\mu_i)=1-\mu_i^2=0$ where $\mu_i=\pm 1$ represents fully left or right circular optical polarization. With the constraints, the corresponding Lagrange function is:
\begin{equation*}
    L(\bm{\mu},\bm{\gamma})=\sum_i\alpha_i\mu_i^2+\sum_{i,j}J_{ij}\mu_i\mu_j-\sum_i\gamma_i\lp\mu_i^2-1\rp.
\end{equation*}
Once again, the gains $\gamma_i$ play the role of Lagrange multipliers. Thus, a minimization of the power dissipation, subject to the optical gain $\gamma_i$, solves the Ising problem defined by the same $J_{ij}$ couplings.

The power dissipation function and constraint function in the Lagrange function above are:
\begin{gather*}
    f(\bm{\mu})=\sum_i\alpha_i\mu_i^2+\sum_{i,j}J_{ij}\mu_i\mu_j,\\
    g_i(\mu_i)=\lp1-\mu_i^2\rp=0.
\end{gather*}

\subsubsection{Conclusions for Coupled Multicore Fibers}
\begin{enumerate}
    \item The coupled multicore fiber system \cite{babaeian_single_2019} minimizes the entropy production rate, or power dissipation, which includes the power input from the pump gain, as in Sections~\ref{yamamoto} and \ref{patrick}.
    \item The coupled multicore fiber system actually performs Lagrange Multiplier optimization.
    \item The gain $\gamma_i$ in each fiber $i$ plays the role of the corresponding Lagrange Multiplier.
    \item Under the constraint of completely light polarization, $\mu_i=\pm 1$, the entropy generation minimization procedure is actually binary, similar to a flip-flop.
\end{enumerate}

\subsection{Coupled Electrical Oscillators on the Unit Circle}
\label{roychow}
\subsubsection{Overview}
In this section, we consider a network of interconnected, nonlinear, but amplitude-stable electrical oscillators designed by Roychowdhury et al. \cite{mcquillan_oim_2019} to represent a conventional Ising system for which we seek a digital solution with each magnetic dipole $\mu_{iz}=\pm 1$ along the z-axis in the magnetic dipole space. To solve this, Roychowdhury et al. provide a dissipative system of LC oscillators, somewhat similar to the Optical Parametric Oscillators in Section~\ref{yamamoto}, but with oscillation amplitude clamped, and oscillation phase $\phi_i=0$ or $\pi$ revealing the preferred magnetic dipole orientation $\mu_{iz}=\pm 1$. It is noteworthy that Roychowdhury goes beyond Ising machines and constructs general digital logic gates using these amplitude-stable oscillators in \cite{roychowdhury_boolean_2015}.

In the construction of their oscillators, Roychowdhury et al. \cite{mcquillan_oim_2019} use nonlinear elements that behave like negative resistors at low voltage amplitudes but with saturating resistance at high voltage amplitudes. This produces of amplitude-stable oscillators. In addition, Roychowdhury et al. \cite{mcquillan_oim_2019} provide a second harmonic pump
and use a form of parametric amplification (referred to as sub-harmonic injection locking in \cite{mcquillan_oim_2019}) to obtain bistability with respect to phase.

The dynamics of the amplitude saturation is purposely very fast and the oscillators are essentially clamped to an amplitude limit dictated by the nonlinear resistor in each oscillator. The phase dynamics induced by the 2\textsuperscript{nd} harmonic pump are slower. It is the readout of these phase shifts, 0 or $\pi$ that provides the magnetic dipole orientation $\mu_{iz}=\pm 1$. One key difference between this system and Yamamoto’s Optical Parametric Oscillator (OPO) system is that the OPO system had fast phase dynamics and slow amplitude dynamics, while the injection locking system has the reverse.

\subsubsection{Equations and comparison with Lagrange Multipliers}
Roychowdhury et al. \cite{mcquillan_oim_2019} derived the dynamics of their amplitude stable oscillator network using perturbation concepts developed in \cite{demir_phase_2000}. While a circuit diagram is not shown, \cite{mcquillan_oim_2019} invokes the following dynamical equation for the phases of their electrical oscillators:
\begin{equation}
    \frac{d\phi_i}{dt}=-\frac{1}{R_c}\sum_{j}{J_{ij}\sin{\lp\phi_i(t)-\phi_j(t)\rp}}-\lambda_i\sin{\lp2\phi_i(t)\rp}, \label{11}
\end{equation}
where $R_c$ is a coupling resistance in their system, $\phi_i$ is the phase of the $i$-th electrical oscillator, and the $\lambda_i$ are decay parameters that dictate how fast the phase angles settle towards their steady state values.

We shall now show that Eq \eqref{11} can be reproduced by iteratively minimizing the power dissipation in their system. Power dissipation across a resistor is $(V_1-V_2)^2/R_c$ where $(V_1-V_2)$ is the voltage difference across resistor $R_c$. Since $V_1$ and $V_2$ are sinusoidal, the power dissipation consists of constant terms and a cross-term of the form:
\begin{equation*}
    f(\phi_1,\phi_2)=\frac{|V|^2\cos{\lp\phi_1-\phi_2\rp}}{R_c},
\end{equation*}
where $f(\phi_1,\phi_2)$ is the power dissipated in the resistors. Magnetic dipole orientation parallel or anti-parallel is represented by whether electrical oscillator phase difference is $\phi_1-\phi_2=0$ or $\phi_1-\phi_2=\pi$ respectively. We are permitted to choose an origin for angle space at $\phi=0$ which implies $\phi_i=0$ or $\phi_i=\pi$. This can be implemented through a constraint of the form:
\begin{equation*}
    g_i(\phi_i)=\lp\cos{\lp2\phi_i\rp}-1\rp=0.
\end{equation*}
Combining the power dissipated in the resistors with the constraint function $g_i(\phi_i)=0$ we obtain a Lagrange function:
\begin{equation*}
    L(\bm{\phi},\bm{\lambda})=\frac{1}{R_c}\sum_{i,j}{J_{ij}\cos{\lp\phi_i-\phi_j\rp}}+\sum_i{\lambda_i\lp\cos{\lp2\phi_i\rp}-1\rp},
\end{equation*}
where $\lambda_i$ is the Lagrange Multiplier corresponding to the phase angle constraint, and $J_{ij}$ is a digital multiplier $\pm 1$ to the conductance $\frac{1}{R_c}$.

The Lagrange function above is isomorphic with the general form in Section~\ref{Lagrange tut}. The effective merit function $f$ and constraints $g_i$ in this correspondence are:
\begin{gather*}
    f(\bm{\phi})=\frac{1}{R_c}\sum_{i,j}{J_{ij}\cos{\lp\phi_i-\phi_j\rp}},\\
    g_i(\phi_i)=\lp\cos{\lp2\phi_i\rp}-1\rp=0.
\end{gather*}

\subsubsection{Conclusions from Coupled Oscillators on the Unit Circle}
\begin{enumerate}
    \item As in other cases, the amplitude-stable oscillator system \cite{mcquillan_oim_2019} minimizes the entropy production rate, or power dissipation, subject to constraints of amplitude and phase reference.
    \item The amplitude-stable oscillator system actually performs Lagrange Multiplier optimization.
    \item The phase decay time constant $\lambda_i$ in each oscillator $i$ plays the role of the corresponding Lagrange Multiplier.
    \item Under the phase reference constraint, $\phi_i=0$ or $\pi$, the entropy generation minimization scheme is actually binary, similar to a flip-flop.
\end{enumerate}

\subsection{Coupled polariton condensates}
\label{cambridge}
\subsubsection{Overview}
Kalinin and Berloff \cite{kalinin_global_2018} proposed a system consisting of coupled polariton condensates to minimize the XY Hamiltonian. The XY Hamiltonian is a 2 dimensional version of the Ising Hamiltonian and is given by:
\begin{equation*}
    H(\bm{\mu})=\sum_{ij}J_{ij}\bm{\mu_i}\cdot\bm{\mu_j},
\end{equation*}
where the $\bm{\mu_i}$ represents the magnetic moment vector of the $i$-th spin restricted to the spin-space XY plane.

To represent the spin system, Kalinin and Berloff pump a grid of coupled semiconductor microcavities with laser beams and observe the formation of strongly coupled exciton-photon states called polaritons. For our purposes, the polaritonic nomenclature is irrelevant. For us, these are simply coupled electromagnetic cavities similar to optical and LC resonators that we have already discussed. The electromagnetic system operates by the principle of minimum entropy generation similar to the previous cases. The complex electromagnetic amplitude in the $i$-th microcavity can be written $E_i = c_i+\text{j}s_i$, where $c_i$ and $s_i$ represent the cosine and sine quadrature components of $E_i$, and j is the unit imaginary. In this case we identify $\text{Re}(E)=c$ as representing the X-component of the magnetic dipole vector, and $\text{Im}(E)=s$ representing the Y-component of the magnetic dipole vector. The electromagnetic microcavity system settles into a state of minimum entropy generation as the laser pump and optical gain are ramped up to compensate for the intrinsic cavity losses. The phase angles in the complex plane of the final electromagnetic modes are then reported as the corresponding $\bm{\mu}$-magnetic moment angles in the XY plane.

An important point to note here is that the electromagnetic cavities experience normal phase-independent gain and not parametric gain which happens to be phase-dependent. As a consequence, this system does NOT seek phase bistability as appropriate for binary up/down spin orientations. This is because we are searching for the magnetic dipole vector angles in the XY plane which would minimize the corresponding XY magnetic energy.

\subsubsection{Lagrange function and Lagrange multipliers}

Ref. \cite{kalinin_global_2018} uses `Ginzburg-Landau' equations to analyze their system resulting in equations for the complex amplitudes $\Psi_i$ of the polariton wavefunctions. But the $\Psi_i$ are actually the complex electric field amplitudes $E_i$ of the corresponding $i$-th cavity. The electric field amplitudes satisfy the following slowly-varying-amplitude equation:
\begin{equation}
    \frac{dE_i}{dt}=\lp\gamma_i-\beta|E_i|^2\rp E_i-iU|E_i|^2E_i-\sum_j{J_{ij}E_j}, \label{12}
\end{equation}
where $\gamma_i$ represents the optical gain, $\beta$ represents nonlinear attenuation, $U$ represents nonlinear phase shift, and $J_{ij}$ is a dissipative cross-coupling-term representing linear loss. We note from the above that both the amplitudes and phases of the electromagnetic modes are coupled to each other and evolve on comparable timescales. This is in contrast to ref. \cite{mcquillan_oim_2019} where the main dynamics were embedded in phase---amplitude was fast and almost fixed---or conversely \cite{xiao_optoelectronics_2019} embedded in amplitude---phase was fast and almost fixed.

We shall now show that the method of ref. \cite{kalinin_global_2018} is essentially the method of Lagrange multipliers with an added `rotation’. The entropy generation or power dissipation rate is:
\begin{align*}
    h(\bm{E})=&\frac{d}{dt}\sum_i{\lp\frac{E_i^*}{2}+\frac{E_i}{2}\rp^2}\\
    =&\frac{1}{2}\sum_{i,j}{J_{ij}\lp E_i^*E_j+E_iE_j^*\rp}+\sum_i{\beta|E_i|^4}-\sum_i{\gamma_i|E_i|^2}.
\end{align*}
If we add a saturation constraint, $g_i(E_i)=\nobreak\lp1-|E_i|^2\rp=\nobreak0$, then by analogy to the previous sections, $\gamma_i$ is reinterpreted as a Lagrange Multiplier:
\begin{equation}
\begin{split}
    L(\bm{E},\bm{\gamma})=&\frac{1}{2}\sum_{i,j}{J_{ij}\lp E_i^*E_j+E_iE_j^*\rp}+\sum_i{\beta|E_i|^4}\\
    &-\sum_i{\gamma_i\lp|E_i|^2-1\rp}, \label{13}
\end{split}
\end{equation}
where $L$ is the Lagrange function that represents power dissipation combined with the amplitude constraint $|E_i|^2=1$. Thus, the scheme of coupled polaritonic resonators operates to find the state of minimum entropy generation in steady-state, similar to the parametric oscillator case of Section~\ref{yamamoto}. That is, the steady-state solution of dynamics \eqref{12} renders \eqref{13} stationary with respect to changes in $\bm{E}$. The difference is that the coupled polaritonic system solves the XY Ising problem for a magnetic moment restricted to the magnetic XY plane, while the parametric oscillator system, in Section~\ref{yamamoto}, solves the $\pm$Z Ising problem. For any particular mathematical optimization that we would perform, we still retain the burden of specifying that dissipative system whose entropy generation matches the optimization that we are seeking.

The imaginary ‘rotation’ term, $iU$, could potentially be of use in developing more sophisticated algorithms than the method of Lagrange multipliers and we discuss this prospect in some detail in Section~\ref{leleu} where a system with a similar, but more general, ‘rotation’ term is discussed.

\subsubsection{Iterative Evolution of Lagrange Multipliers}
Next, we discuss how Kalinin and Berloff et al. \cite{kalinin_global_2018}, adjusted their Lagrange multipliers $\gamma_i$ during the course of their optimization. In the method of Lagrange multipliers, the merit-function, eq \eqref{13}, is used to optimize not only the electric field amplitudes $E_i$ but also the Lagrange Multipliers $\gamma_i$. The authors in Sections~\ref{yamamoto} and \ref{patrick} used simple heuristics to adjust their gains/decay constants which we have proven to be the Lagrange multipliers. Kalinin and Berloff of this section employ the Lagrange function optimization itself to adjust the gain/losses, as in the complete Lagrange method discussed next.

To iteratively adjust the Lagrange multipliers, we shall briefly shift back to the tutorial notation of Section~\ref{Lagrange tut}.  Afterward, we shall translate it back to the notation of this section and show that Kalinin, Berloff et al., indeed use this same iterative procedure to adjust their Lagrange multipliers. 

The Lagrange function is given by $L(\bm{x},\bm{\lambda})=f(\bm{x})+\sum_{i=1}^p{\lambda_ig_i(\bm{x})}$, where $x_i$ are the field variables and $\lambda_i$ are the Lagrange multipliers. Then, the procedure to find the optimal $\bm{x^*}$ and $\bm{\lambda^*}$ is to perform gradient descent of $L$ in $\bm{x}$ and gradient ascent of $L$ in $\bm{\lambda}$. The reason for ascent in $\bm{\lambda}$ rather than descent is to more strictly penalize deviations from the constraint. In the language of iterations, this leads to the expressions:
\begin{gather*}
    x_i(t+\Delta t)=x_i(t)-\kappa\Delta t\frac{\partial}{\partial x_i}L(\bm{x},\bm{\lambda}),\\
    \lambda_i(t+\Delta t)=\lambda_i(t)+\kappa'\Delta t\frac{\partial}{\partial \lambda_i}L(\bm{x},\bm{\lambda}),
\end{gather*}
where $\kappa$ and $\kappa'$ are suitably chosen step sizes.

With our identification that the Lagrange multipliers $\lambda$ are actually the same as the gains $\gamma$, we plug the expression for the full Lagrange function \eqref{13} into the second iterative equation, projecting out the constraint function $g_i(E_i)=\lp1-|E_i|^2\rp=0$, and take the limit $\Delta t\to 0$. We obtain the following dynamical equation for the gains $\gamma_i$:
\begin{equation}
    \frac{d\gamma_i}{dt}=\kappa'\lp1-|E_i|^2\rp. \label{14}
\end{equation}
The above equation for the iterative evolution of the Lagrange multipliers is indeed the very same evolution that Kalinin, Berloff et al. employ in their coupled polariton system.

To Eq \eqref{14}, we must add the iterative evolution of the field variables $x_i$:
\begin{equation}
    \frac{dx_i}{dt}=-\kappa\frac{\partial}{\partial x_i}L(\bm{x},\bm{\lambda}). \label{15}
\end{equation}
Equations \eqref{14} and \eqref{15} represent the full iterative evolution, but in some of the earlier sub-sections, $\gamma_i(t)$ was sometimes assigned a heuristic time dependence.

We conclude this sub-section by splitting the Lagrange function into the effective merit function $f$, and the constraint function $g_i$. The extra `phase rotation' U is not captured by this interpretation.
\begin{gather*}
    f(E_1,\dots,E_n)=\frac{1}{2}\sum_{i,j}{J_{ij}\lp E_i^*E_j+E_iE_j^*\rp}+\sum_i{\beta|E_i|^4},\\
    g_i(E_i)=\lp1-|E_i|^2\rp=0.
\end{gather*}
\subsubsection{Conclusions from Coupled Electromagnetic Cavities}
\begin{enumerate}
    \item As in other cases, the coupled-cavity system \cite{kalinin_global_2018} minimizes the steady state entropy production rate, or power dissipation, which includes the power input from the pump gain.
    \item The gain $\gamma_i$ in each oscillator $i$ plays the role of the corresponding Lagrange Multiplier.
    \item There is a phase rotation term $iU$ in the dynamics that is not captured by the Lagrange function framework.
\end{enumerate}

\section{Other methods in the literature}
\label{other}
In this section, we look at other methods in the literature that do not explicitly implement the method of Lagrange Multiplier but nevertheless end up with dynamics that resemble it to varying extents. All these methods offer operation regimes where the dynamics is not analogous to Lagrange multiplier optimization, and we believe it is an interesting avenue of future work to study the capabilities of these regimes.   

\subsection{Iterative Analog Matrix Multipliers}
\label{mit}
Soljacic et al. \cite{roques-carmes_heuristic_2020} developed an iterative procedure consisting of repeated matrix multiplication to solve the Ising problem. Their algorithm was implemented on a photonic circuit that utilized on-chip optical matrix multiplication units composed of Mach-Zehnder interferometers that were first introduced for matrix algebra by Zeilinger et al. in \cite{reck_experimental_1994}. Soljacic et al. showed that their algorithm performed optimization on an effective merit function, such as total magnetic Ising energy.

Let us use our insights from the previous sections and see how one would implement iterative optimization using an optical matrix multiplier. Let the multiple magnetic moment configuration of the Ising problem be represented as a vector of electric field amplitudes, $E_i$, of the spatially-separated optical modes. Each mode field amplitude represents the value of each magnetic moment. In each iteration, the optical modes are fed into the optical circuit which performs matrix multiplication, and the resulting output optical modes are then fed back to the optical circuit input for the next iteration. Optical gain or some other type of gain sustains the successive iterations.

\begin{figure}[h]
\begin{center}
\includegraphics[scale=0.5]{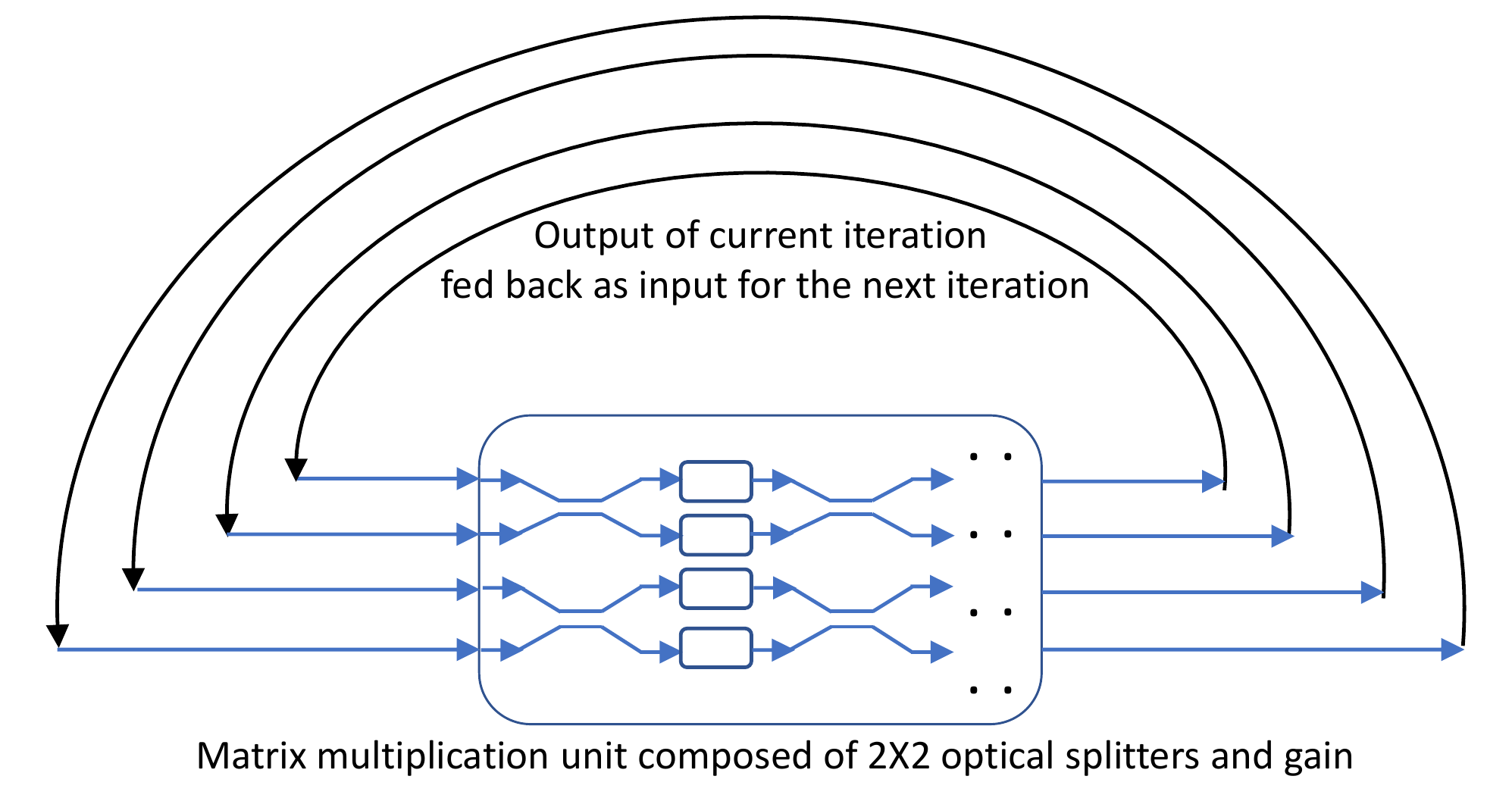}  
\end{center}
% Here is how to import EPS art
\caption{\footnotesize \label{fig:mit} An optical circuit performing iterative multiplications converges on a solution of the Ising problem. Optical pulses are fed as input from the left hand side at the beginning of each iteration, pass through the matrix multiplication unit, and are passed back from the outputs to the inputs for the next iteration. Distributed optical gain sustains the iterations.}
\end{figure}

In this section we design an iterative optimization scheme for the Ising problem that involves only matrix multiplications in each iterative step and whose power dissipation function matches the magnetic Ising energy. A simple block diagram of such a scheme is shown in Fig.~\ref{fig:mit}.
    
We wish to design the matrix multiplication unit such that it has the following power dissipation function:
\begin{equation*}
    h(\bm{E})=-\sum_i{\gamma_i|E_i|^2}+\frac{1}{2}\sum_{i,j}{J_{ij}\lp E_i^*E_j+E_iE_j^*\rp}.
\end{equation*}
The Lagrange function, including a binary constraint, $|E_i|^2=1$, is given by:
\begin{equation}
    L(\bm{E},\bm{\gamma})=-\sum_i{\gamma_i\lp|E_i|^2-1\rp}+\frac{1}{2}\sum_{i,j}{J_{ij}\lp E_i^*E_j+E_iE_j^*\rp}, \label{16}
\end{equation}
where the $J_{ij}$ represents dissipative loss associated with electric field interference between optical modes in the Mach-Zehnder interferometers. The dissipation is compensated by optical gain $\gamma_i$ in the system.

The iterative multiplicative procedure that evolves the electric fields toward the minimum of the Lagrange function Eq \eqref{16} is given by:
\begin{equation*}
\begin{split}
    E_i(t+1)-E_i(t)&=-\kappa\Delta t\frac{\partial}{\partial E_i}\lp\sum_i{\gamma_i\lp1-|E_i|^2(t)\rp}\right.\\
    &+\left.\frac{1}{2}\sum_{i,j}{J_{ij}\lp E_i^*(t)E_j(t)+E_i(t)E_j^*(t)\rp}\rp,
\end{split}
\end{equation*}
where we move from iteration $t$ to iteration $t+1$ by taking steps in $E_i$ proportional to the gradient $\partial/\partial E_i$ of the Lagrange function. ($\partial/\partial E_i$ represents separate differentiation with respect to the two quadrature components.) With each iteration, we feed the output of the Mach-Zehnder interferometer matrix-multiplier array back to the input, compensating for the losses with gain. Calculating the gradient, we obtain:
\begin{equation*}
    E_i(t+1)-E_i(t)=2\kappa\Delta t\lp\gamma_iE_i(t)-\sum_j{J_{ij}E_j(t)}\rp,
\end{equation*}
where $\kappa$ is a constant step size, whose dimensionality should ensure consistency of units. Sending all the terms involving time step $t$ to the right hand side, we finally get:
\begin{equation}
    E_i(t+1)=\sum_j{\lp\lp1+2\kappa\Delta t\gamma_i\rp\delta_{ij}-2\kappa\Delta tJ_{ij}\rp E_j(t)}, \label{17}
\end{equation}
where $\delta_{ij}$ is the Kronecker delta that is 1 only if $i=j$. The Mach-Zehnder interferometers should be tuned to the matrix $\left[\lp1+2\kappa\Delta t\gamma_i\rp\delta_{ij}-2\kappa\Delta tJ_{ij}\right]$. Thus, we have an iterative matrix multiplier scheme that minimizes the Lagrange function and performs Lagrange multiplier optimization of the Ising problem. In effect, a lump of dissipative optical circuitry, compensated by optical gain, will, in a series of iterations, settle into a solution of the Ising problem.

The simple system above differs from that of Soljacic et al. \cite{roques-carmes_heuristic_2020} quite significantly, in that their method has added noise and nonlinear thresholding after each iteration. It is possible that their modifications lead to performance improvements, or possibly it might work equally well as our more standard approach. A detailed description of the Soljacic approach is presented in Appendix E.

\subsection{Leleu Mathematical Ising Solver}
\label{leleu}
Leleu et al. \cite{leleu_destabilization_2019} proposed a modified version of the Yamamoto’s Ising machine \cite{haribara_computational_2016}. Leleu’s method significantly resembles the Lagrange method while incorporating important new features that might be important research directions in their own right. To understand the similarities and differences between Leleu’s method and that of Lagrange multipliers, we shall first list out the Lagrange multipliers equations of motion.

We recall the Lagrange function for the Ising problem that we encountered in Section~\ref{Ising solvers}:
\begin{equation*}
    L(\bm{x},\bm{\gamma})=\sum_{i,j}{J_{ij}x_ix_j}+\sum_i{\alpha_ix_i^2}+\sum_i{\gamma_i\lp1-x_i^2\rp}.
\end{equation*}
In the above, $x_i$ are the optimization variables, $J_{ij}$ is the interaction matrix, $\gamma_i$ is the gain provided to the $i$-th variable, and $\alpha_i$ is the loss experienced by the $i$-th variable. To find a local optimum $(\bm{x^*},\bm{\gamma^*})$ that satisfies the constraints, one has to perform gradient descent on the Lagrange function in the $\bm{x}$ variables and gradient ascent in the $\bm{\gamma}$ variables as discussed in Section~\ref{cambridge}. That is, the iterations one would need to perform are:
\begin{gather*}
    x_i(t+\Delta t)=x_i(t)-\kappa\Delta t\frac{\partial}{\partial x_i}L(\bm{x},\bm{\gamma}),\\
    \gamma_i(t+\Delta t)=\gamma_i(t)+\kappa'\Delta t\frac{\partial}{\partial \gamma_i}L(\bm{x},\bm{\gamma}),
\end{gather*}
where $\kappa$ and $\kappa'$ are suitably chosen step sizes. Substituting the expression for $L$ into the above and taking the limit of $\Delta t\to 0$, we get the following equations of motion:
\begin{gather}
    \frac{dx_i}{dt}=2\kappa\lp\lp-\alpha_i+\gamma_i\rp x_i-\sum_j{J_{ij}x_j}\rp, \label{18}\\
    \frac{d\gamma_i}{dt}=\kappa'\lp1-x_i^2\rp. \label{19}
\end{gather}
We shall compare these equations with those designed by Leleu et al. in \cite{leleu_destabilization_2019}. Their system is described by the following equations:
\begin{gather}
    \frac{dx_i}{dt}=(-\alpha+\gamma)x_i+e_i\sum_j{J_{ij}x_j}, \label{20}\\
    \frac{de_i}{dt}=\beta(1-x_i^2)e_i, \label{21}
\end{gather}
where the $x_i$ represent the optimization variables as usual, $\alpha$ represents the loss experienced by each degree of freedom, $\gamma$ represents a common gain supplied to each variable, $\beta$ is some suitably chosen positive parameter, and the $e_i$ represent error coefficients that capture how far away each $x_i$ is from its desired unity saturation amplitude. Leleu had cubic terms in $x_i$ in \cite{leleu_destabilization_2019} but we shall ignore them in the present discussion for the sake of simplicity. A discussion of these cubic terms is given in Appendix C.  

It is clear at once that there are significant similarities between Leleu’s system and the Lagrange multiplier system. The optimization variables in both systems experience linear losses and gains and have interaction terms that capture the Ising interaction. Both systems have auxiliary variables that are varied according to how far away each degree of freedom is from its preferred saturation amplitude. However, the similarities end here.

A major differentiation in Leleu’s system is that $e_i$ multiplies the Ising interaction felt by the $i$-th variable resulting in $e_iJ_{ij}$. On the other hand, the complementary coefficient is $e_jJ_{ij}$. This has the consequence that Leleu’s equations implements a system which has non-symmetric interactions $e_iJ_{ij} \ne e_jJ_{ij}$ between vector components $x_i$ and $x_j$. The inclusion of non-symmetric matrix terms seems to be important because Leleu’s system achieves excellent performance on Ising problems in the Gset problem set as demonstrated in \cite{leleu_destabilization_2019}.

Let us obtain some intuition about this system by splitting the non-symmetric interaction term $e_iJ_{ij}$ into the sum of a symmetric and anti-symmetric part. This follows from the fact that any matrix $A$ can be written as the sum of a symmetric matrix, $(A+A^T)/2$, and an anti-symmetric matrix, $(A-A^T)/2$. The symmetric part leads to gradient descent dynamics just as in Eq \eqref{20}, similar to all the systems in Section~\ref{Ising solvers}. Conversely, the anti-symmetric part causes a energy-conserving ‘rotary’ motion in the vector space of $x_i$. Leleu et al.’s non-symmetric approach seems to contain the same number of auxiliary parameters, $e_i$ versus $\gamma_i$, and so the secret of their improved performance seems to lie in this anti-symmetric part. We believe further analysis of this method might be a fruitful future research direction.

We recall Onsager’s reciprocity theorem \cite{onsager_reciprocal_1931} which states that the thermodynamic response coefficient $R_{ij}$ must be equal to the response coefficient $R_{ji}$ if time-reversal symmetry holds in the equilibrium state. This would imply that one might have to construct a magnetic system of some sort to break time-reversal symmetry in order to physically implement the asymmetry embedded in the dynamics designed by Leleu et al.

In conclusion, the symmetric part of Eq \eqref{20} takes iterative steps along the gradient of a dissipative merit function. The anti-symmetric part produce energy conserving rotations in the the vector space of optimization variables, $x_i$. The associated dynamical freedom might provide a fruitful future research direction in optimization and deserves further study to ascertain its power.

%%%%%%%%%%%%%%%%%%%%%%%%%%%%%%%%%%%%%%%%%%%%%%%%%%%%%%%%%%%%%%%%%%%%%%%%%%%%%%%%%%%%%%%%%%%%%%%%%%

\section{Applications in Linear Algebra and Statistics}
\label{reg}
We have seen that minimum power dissipation solvers can address the Ising problem and various similar problems like the traveling salesman problem, etc. In this section, we provide yet another application of minimum entropy generation solvers to another optimization problem that appears frequently in statistics, namely curve fitting. In particular, we make the observation that the problem of linear least squares regression, linear curve fitting with a quadratic merit function, resembles the Ising problem. In fact, the electrical circuit example we presented in Section~\ref{patrick} can be applied to linear regression. We present such a circuit in this section. Our circuit provides a digital answer but requires a series of binary resistance values, that is, $\dots,2R_0, R_0, 0.5R_0,\dots$ to represent arbitrary binary statistical input observations.

\begin{figure}[h]
\begin{center}
\includegraphics[scale=0.4]{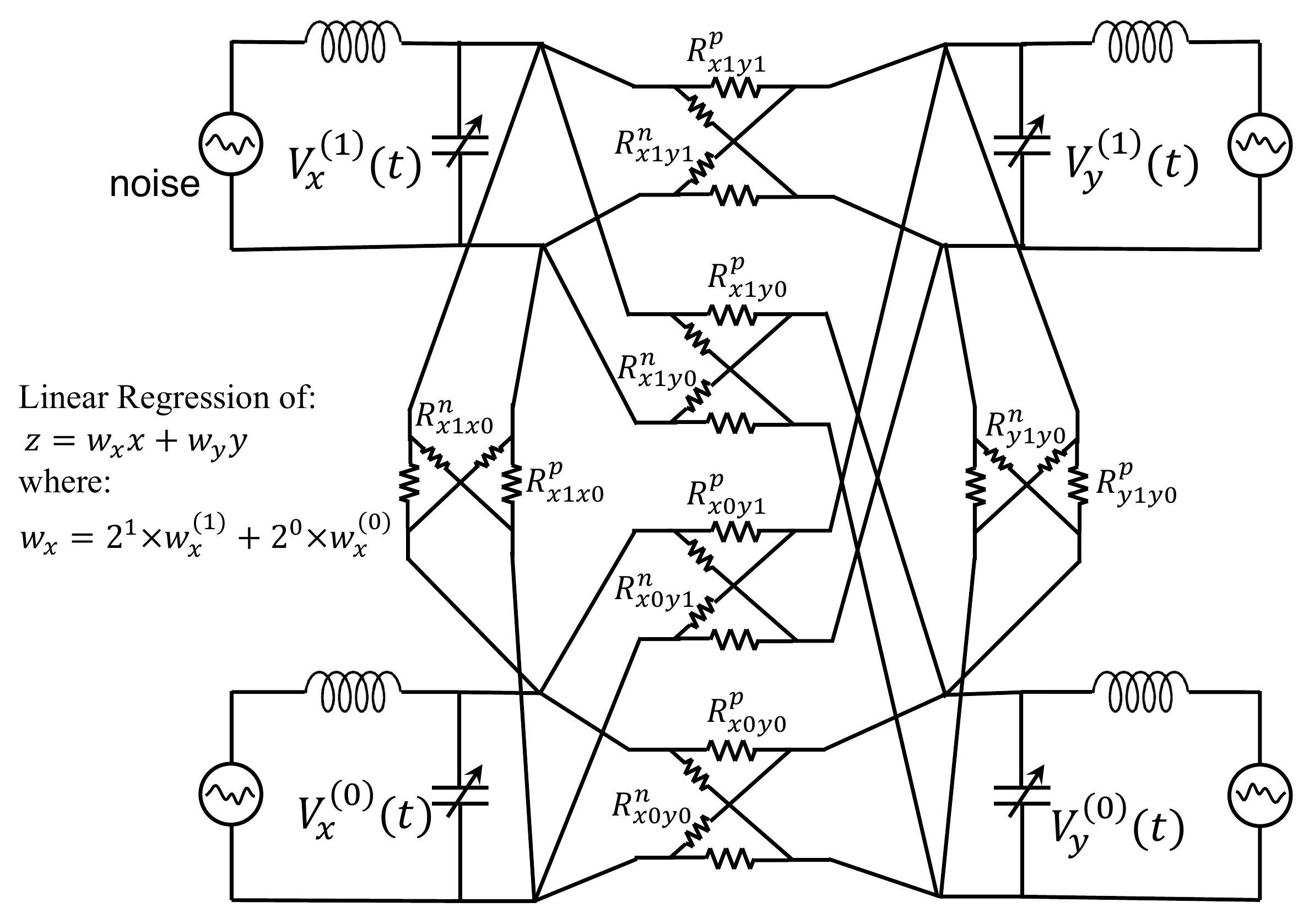}  
\end{center}
% Here is how to import EPS art
\caption{\footnotesize \label{fig:reg} A two-bit linear regression circuit to find the best two curve-fitting weights $w_d$ using the Principle of Minimum Entropy Generation.}
\end{figure}

The objective of linear least squares regression is to fit a linear function to a given set of data $\{(\bm{x_1}, y_1), (\bm{x_2}, y_2), (\bm{x_3}, y_3),\dots, (\bm{x_n}, y_n)\}$. The $\bm{x_i}$ s are input vectors of dimension $d$ while the $y_i$ are the observed outputs that we want our regression to capture. The linear function that is being fit is of the form $y(\bm{a})=\sum_{i=1}^d{w_ia_i}$ where $\bm{a}$ is a feature vector of length $d$ and $\bm{w}$ is a vector of unknown weights multiplying the features vector $\bm{a}$. The vector $\bm{w}$ is calculated by minimizing the sum of the squared errors it causes when used on an actual data set. This optimization problem may be represented as:
\begin{equation*}
    \bm{w^*}=\min_{\bm{w}}\sum_{i=1}^n{\left[\lp\sum_{j=1}^d{w_jx_{ij}}\rp-y_i\right]^2},
\end{equation*}
where $x_{ij}$ is the $j$-th component of the vector $\bm{x_i}$. The merit function, upon expansion, yields $\sum_{i=1}^dy_i^2-2\sum_{i=1}^d\lp\sum_{j=1}^nx_{ji}y_j\rp w_i+\sum_{i,j=1}^d\lp\sum_{k=1}^nx_{ki}x_{kj}\rp w_iw_j$. This functional form is identical to that of the Ising Hamiltonian and we may construct an Ising circuit with $J_{ij}=\sum_{k=1}^nx_{ki}x_{kj}$ with the weights $\bm{w}$ acting like the unknown
magnetic moments. There is an effective magnetic field in the problem, $h_i=-2\sum_{j=1}^nx_{ji}y_j$, and a fixed number $\sum_{i=1}^dy_i^2$ which doesn’t play a role in the optimization. A simple circuit that solves the linear least squares problem for $d=2$---the case where there are two features per instance and, consequently, two weights $w_d$ to be estimated---is provided in Fig.~\ref{fig:reg}. This circuit provides weights upto 2-bit precision.

The two oscillators on the left hand side of the figure represent the $2^0$ and $2^1$ bits of the weight corresponding to the first feature while the oscillators on the right hand side are the corresponding bits of the second weight.

The cross-resistance $R$ that one would need to represent the $J_{ij}$ that connects the $i$-th and $j$-th oscillators is calculated as:
\begin{equation*}
    \frac{1}{R}=\frac{b_1}{R_{-1}}+\frac{b_0}{R_{0}}+\frac{b_{-1}}{R_{1}},
\end{equation*}
where $R_m=2^mR_0$ is a binary hierarchy of resistances $\dots,2R_0, R_0, 0.5R_0,\dots$ based on a reference resistor $R_0$, and $b_m$ are the binary digits of $J_{ij}$ expressed as a binary number: $b=b_1\times 2^1+b_0\times 2^0+b_{-1}\times 2^{-1}$. This represents $J_{ij}$ to 3-bit precision using resistors that span a dynamic range $2^2=4$. Further, the sign of the coupling is allotted according to whether the resistors $R$ are parallel-connected or cross-connected. In operation, the resistors $R$ would be externally programmed to the correct binary values, with many more bits than 3-bit precision, as given by the matrix product $J_{ij}=\sum_{k=1}^nx_{ki}x_{kj}$.

We have just solved the regression problem of the form $\bm{Xw}=\bm{y}$, where matrix $\bm{X}$ and vector $\bm{y}$ were known measurements and the corresponding best weight vector $\bm{w}$ for fitting was the unknown. We conclude by noting that this same procedure can be adopted to solve linear systems of equations of the form $\bm{Xw}=\bm{y}$.

%%%%%%%%%%%%%%%%% Section 4 %%%%%%%%%%%%%%%%%%%%%%%%%%%%%%%%%%%%%%%%%%%%%%%%%%%%%%

\section{Discussion and Conclusion}
\label{conc}
Physics obeys a number of maximization or minimization principles such as the principle of Least Action, the principle of Minimum Entropy Generation, the Variational Principle, Physical Annealing, and the Adiabatic Principle (which in its quantum form is called Quantum Annealing).

Optimization is very significant in diverse fields ranging from scheduling and routing in operations research to protein folding in biology, porfolio optimization in finance, and energy minimization in physics. In this article, we made the observation that physics has optimization principles at its heart, and that they can be exploited to design fast, low power, digital solvers that avoid the limits of standard computational paradigms. Nature thus provides us with a means to solve optimization problems in all these areas including Engineering, Artificial Intelligence, Machine Learning (backpropagation), Control Theory, and Reinforcement Learning.

We reviewed 7 physical machines, proposed or built, that purported to solve the Ising problem and found that 6 of the 7 were performing Lagrange multiplier optimization; further, they also obey the Principle of Minimized Entropy generation (always subject to a power input constraint). This means that by appropriate choice of parameter values, these physical solvers can be used to perform Lagrange multiplier optimization orders-of-magnitude faster and with lower power than conventional digital computers. This performance advantage can be utilized for optimization in machine learning applications where energy and time considerations are critical.

The questions arise: What are the action items? And what is the most promising near term application? All the hardware approaches seem to work comparably well. The easiest to implement would be the electrical oscillator circuits, though the optical oscillator arrays can be compact and very fast. Electrically, there would two integrated circuits, the oscillator array, and the connecting resistors that would need to be reprogrammed for different problems. The action item could be to design the first chip consisting of about 1000 oscillators, and a second chip that would consist of the appropriate coupling resistor array, for a specific optimization problem. The resistors should be in an addressable binary hierarchy so that any desired resistance value can be programmed in by switches, within the number of bits accuracy. It is possible to imagine solving a new Ising problem ever milli-second, by reprogramming the resistor chip.

On the software side, a compiler would need to be developed to go from an unsolved optimization problem to the resistor array which matches that desired goal. If the merit function were mildly nonlinear, we believe that the Principle of Minimum Entropy generation would still hold, but there has been less background science justifying that claim.

With regard to the most promising near term application, it might be in Control Theory or in Reinforcement Learning in self-driving vehicles, where rapid answers are required, at modest power dissipation.

The act of computation can be regarded as a search among many possible answers. Finally, the circuit converges to a final correct configuration. Thus the initial conditions may include a huge phase space volume=$2^n$ of possible solutions, ultimately transitioning into a final configuration representing a small or modest-sized binary number. This type of computing implies a substantial entropy reduction. This led to Landauer’s admonition that computation costs $kn\log{2}$ of entropy decrease, and $kTn\log{2}$ of energy, for a final answer with $n$ binary digits.

By the 2\textsuperscript{nd} Law of Thermodynamics, such an entropy reduction must be accompanied by an entropy increase elsewhere. For example, in a dissipative circuit, electricity converts to heat. In Landauer’s viewpoint, the energy and entropy limit of computing was associated with the final acting of writing out the answer in $n$-bits, assuming the rest of the computer was reversible. In practice, technology consumes $\sim10^4\times$ times more than the Landauer limit, owing to the insensitivity of the transistors operating at $\sim$1Volt, when they could be operating at $\sim$10mVolts.

In the continuously dissipative circuits we have described here, the energy consumed would be infinite if we waited long enough for the system to reach the final optimal state. If we terminate the powering of our optimizer systems after they
reach the desired final state answer, the energy consumed becomes finite. By operating at voltage $<$1Volt and by powering off after the desired answer is achieved, our continuously dissipating Lagrange optimizers could actually be closer to the Landauer limit than a conventional computer.

A controversial point relates to the quality of solutions that are obtained for \textbf{NP}-hard problems. The physical systems we are proposing evolve by steepest descent toward a local optimum, not a global optimum. Nonetheless, many of the authors of the 7 physical systems presented here have claimed to find better local optima than their competitors, due to special adjustments in their methods. Undoubtedly, some improvements are possible, but none of the 7 papers reviewed here claims to always find the one global optimum which would be \textbf{NP}-hard \cite{karp_reducibility_1972}.

We have shown that a number of physical systems that perform optimization are acting through the Principle of Minimum Entropy generation, though other physics principles could also fulfill this goal. As the systems evolve toward an extremum, they perform Lagrange function optimization where the Lagrange Multipliers are given by the gain or loss coefficients that keep the machine running. Thus, Nature provides us with a series of physical Optimization Machines that are much faster and possibly more energy efficient than conventional computers.

%%%%%%%%%%%%%%%%%%% Acknowledgments %%%%%%%%%%%%%%%%%%%%%%%%%%%%%%%%%%%%%%%%%%%%%%%

The authors gratefully acknowledge useful discussions with Dr. Ryan Hamerly, Dr. Tianshi Wang, and Prof. Jaijeet Roychowdhury. 
The work of S. K. Vadlamani and
Prof. E. Yablonovitch was supported by the National Science Foundation through the
Center for Energy Efficient Electronics Science $(E^3S)$ under Award
ECCS-0939514 and the Office of Naval Research under grant \#N00014-14-1-0505.
 
\appendix

\section{Lagrange multipliers theory}

In this appendix, we shall study in greater detail the method of Lagrange multipliers, one of the most well-known techniques for the solution of constrained optimization problems. At its core, it simply involves the modification of the merit function by adding terms that penalize constraint violations.

In the setting of constrained optimization, we are required to minimize an objective function $f(\bm{x})$ with respect to all points that satisfy certain given inequality ($h_j(\bm{x})\leq 0$) and equality ($g_i(\bm{x})=0$) constraints. Let us assume that the domain set $D$, that is, the intersection of the domains of $f(\bm{x}),g_i(\bm{x}),h_j(\bm{x})$, is a subset of $\mathbb{R}^n$. Further, let the set of points in $D$ that satisfy all the constraints $g_i,h_j$ be called $F$, the feasible set. Then, the optimization problem can be written as:
\begin{alignat*}{2}
    &\text{minimize}\ \ \ &&f(\bm{x})\\
    &\text{subject to }\ \ &&h_j(\bm{x})\leq 0,\ j=1,\dots,m,\\
    &\ &&g_i(\bm{x})=0,\ i=1,\dots,p,\\
    &\ &&\bm{x}\in D.
\end{alignat*}

\subsection{Lagrange function}

We now define a new function in $n+m+p$ variables, called the Lagrange function, as follows:
\begin{equation*}
    L(\bm{x}, \bm{\lambda}, \bm{\mu}) = f(\bm{x}) + \sum_{j=1}^m\mu_jh_j(\bm{x}) + \sum_{i=1}^p\lambda_ig_i(\bm{x}).
\end{equation*}

The summations that were added to the plain objective function $f(\bm{x})$ serve as constraint violation penalty terms. The coefficients multiplying the penalty terms, $\lambda_i$ and $\mu_j$, are known as Lagrange multipliers. The inequality Lagrange multipliers $\mu_j$ are constrained to be non-negative in order that the penalty that arises when the inequality constraints are not satisfied (i.e. when $h_j(\bm{x})> 0$) is non-negative. The equality Lagrange multipliers, $\lambda_i$, have no such restrictions.

The Lagrange function has the advantage that it helps us express the constrained optimization of $f$ as an unconstrained optimization of $L$. That is, it can be shown that:
\begin{equation*}
    \min_{\bm{x}\in F}f(\bm{x})=\min_{\bm{x}\in D}\max_{(\bm{\mu}\geq 0),\bm{\lambda}}L(\bm{x},\bm{\lambda},\bm{\mu}).
\end{equation*}
The minimization over $\bm{x}$ in the feasible set, $F$, on the left-hand side has turned into a minimization over $\bm{x}$ in the entire domain, $D$, on the right-hand side.

\subsection{Karush-Kuhn-Tucker (KKT) sufficient conditions}
The Lagrange function also appears in an important, related role. While the conditions for a point $\bm{x^*}$ to be an unconstrained optimum of a differentiable function $f$ are expressed in terms of the gradient of $f(\bm{x})$, the optimality conditions for constrained optimization problems are naturally expressed in terms of the Lagrange function. These conditions are presented next. 

A point $\bm{x^*}$ is a local optimum of the function $f(\bm{x})$ subject to the constraints $g_i,h_j$ if it satisfies the Karush-Kuhn-Tucker (KKT) sufficient conditions:
\begin{enumerate}
\item Primal feasibility: The point $\bm{x^*}$ is feasible, that is, it satisfies all the constraints.
\begin{align*}
    h_j(\bm{x^*})&\leq 0,\ j=1,\dots,m,\\
    g_i(\bm{x^*})&=0,\ i=1,\dots,p.
\end{align*}
\item First-order condition: There exist Lagrange multipliers $\mu_1^*,\dots,\mu_m^*,\lambda^*_1,\dots,\lambda^*_p$ such that the following equation is satisfied:
\begin{align*}
    &\ \ \ \ \bm{\nabla_x} L(\bm{x^*},\bm{\mu^*},\bm{\lambda^*})\\
    &=\bm{\nabla_x} f(\bm{x^*})+\sum_{j=1}^m\mu_j^*\bm{\nabla_x} h_j(\bm{x^*})+\sum_{i=1}^p\lambda_i^*\bm{\nabla_x} g_i(\bm{x^*})\\
    &=0.
\end{align*}
\item Second-order condition: In addition to the first-order condition, if all the concerned functions are twice differentiable, we require that the Hessian of $L$ with respect to $\bm{x}$ be positive definite along all directions that respect the active constraints. That is, the following equation has to be satisfied:
\begin{align*}
    \bm{v}^T\bm{\nabla_{xx}}^2L(\bm{x^*}, \bm{\mu^*}, \bm{\lambda^*})\bm{v}>0,\ \forall \bm{v}\in T,
\end{align*}
where $T$ is defined as $T=\{\bm{v}:\ \left(\bm{\nabla_x}h_j(\bm{x^*})\right)^T\bm{v}=0,\ j=$ active constraints
at $\bm{x^*},\ \left(\bm{\nabla_x}g_i(\bm{x^*})\right)^T\bm{v}=0,\ i=1,\dots,p\}$.
\item Complementary slackness:  $\mu_j^*h_j(\bm{x^*})=0$ is satisfied for all the inequality constraints, $j=1,\dots,m$. 
\item Dual feasibility: The Lagrange multipliers of all the inequality constraints satisfy $\mu_j^*\geq 0,\ j=1,\dots,m$, with the inequality being strict for active constraints.
\end{enumerate}
We recognize condition number 2 as the one we encountered in Section~\ref{Lagrange tut} in the main text.

\subsection{Saddle point nature of global and local optima}
If the functions $f$, $g_i$, $h_j$, and the feasible set $F$ satisfy certain special conditions (such as, for instance, the Slater condition), a phenomenon called strong duality occurs. Under such circumstances, it turns out that the minimization and maximization in the above equation can be switched:
\begin{equation*}
  \min_{\bm{x}\in D}\max_{(\bm{\mu}\geq 0),\bm{\lambda}}L(\bm{x},\bm{\lambda},\bm{\mu})=\max_{(\bm{\mu}\geq 0),\bm{\lambda}}\min_{\bm{x}\in D}L(\bm{x},\bm{\lambda},\bm{\mu}).  
\end{equation*}
When this occurs, it can be shown that the optimal pair $(\bm{x^*},\bm{\lambda^*},\bm{\mu^*})$ that satisfies the KKT conditions actually forms a saddle point of the Lagrange function, $L(\bm{x},\bm{\lambda},\bm{\mu})$. That is, one has:
\begin{alignat*}{2}
    L(\bm{x^*},\bm{\lambda^*},\bm{\mu^*})&\leq L(\bm{x},\bm{\lambda^*},\bm{\mu^*}),\ &&\forall\ \bm{x}\in D,\\
    L(\bm{x^*},\bm{\lambda^*},\bm{\mu^*})&\geq L(\bm{x^*},\bm{\lambda},\bm{\mu}),\ &&\forall\ \bm{\mu}\geq 0,\ \bm{\lambda}.
\end{alignat*}

This can be viewed as a simple motivation for the gradient \emph{descent} of $L$ in $x$ and gradient \emph{ascent} of $L$ in $\lambda$ that we encountered in Section~\ref{cambridge}. In the next section, Appendix B, we shall follow Chapter 4 of the excellent book \cite{bertsekas_nonlinear_1999} and see how this descent/ascent procedure can be used to find local optima even if strong duality doesn't hold.   

\section{Lagrange multipliers algorithms: Augmented Lagrangian}

In this section, we discuss the so-called `Augmented Lagrangian method of multipliers', a popular algorithm used to obtain locally optimal solutions $(\bm{x^*},\bm{\lambda^*})$ that satisfy the KKT conditions. We shall be only considering the case where there are no inequality constraints $h_j$ (and, consequently, no $\bm{\mu}$ multipliers). This algorithm is discussed in detail in \cite{bertsekas_nonlinear_1999}.

To motivate the `Augmented Lagrangian' approach, let us observe the KKT conditions a bit closely. We conclude from the first-order condition that a locally optimum $\bm{x^*}$ renders the function $L(\bm{x},\bm{\lambda^*})$ stationary. However, this doesn't guarantee that $\bm{x^*}$ is a minimum of $L(\bm{x},\bm{\lambda^*})$. Indeed, observation of the second-order condition tells us that $\bm{x^*}$ could be a saddle point of $L(\bm{x},\bm{\lambda^*})$. This means that gradient descent-based algorithms will not converge to $\bm{x^*}$ if the starting point is not in the correct region. It was to solve this problem that the `Augmented Lagrangian' method was invented.        

The Augmented Lagrange function, $L_c(\bm{x},\bm{\lambda})$, is given by:
\begin{equation*}
    L_c(\bm{x},\bm{\lambda})=f(\bm{x})+\sum_{i=1}^p\lambda_ig_i(\bm{x})+\frac{c}{2}\left(\sum_{i=1}^p\left(g_i(\bm{x})\right)^2\right).
\end{equation*}
This function $L_c(\bm{x},\bm{\lambda})$, for a suitable choice of $c$, has the property that $\bm{x^*}$ forms a minimum of $L_c(\bm{x},\bm{\lambda^*})$ and not just a saddle point as was the case with $L(\bm{x},\bm{\lambda^*})$. 

In other words, local optima of the original constrained optimization problem can be obtained if we perform an unconstrained optimization of $L_c(\bm{x},\bm{\lambda^*})$. However, for this procedure to be a feasible solution approach, we would have to know the right $\bm{\lambda^*}$. It has been shown in \cite{bertsekas_nonlinear_1999} that the way to do this is to perform gradient ascent of $L_c$ in the $\bm{\lambda}$ variables.      

The `Augmented Lagrangian method of multipliers' involves the repeated minimization of $L_c(\bm{x},\bm{\lambda}^{(k)})$ using progressively better estimates, $\bm{\lambda}^{(k)}$, of $\bm{\lambda^*}$. The algorithm starts off with an arbitrary starting point $(x^{(0)}, \lambda^{(0)})$. It then performs the following steps repeatedly:
\begin{enumerate}
    \item Locally minimize $L_c(\bm{x}, \bm{\lambda}^{(k)})$ and call the minimum point $\bm{x}^{(k)}$.
    \item $\lambda_i^{(k+1)}=\lambda_i^{(k)}+cg_i(\bm{x}^{(k)})$.
\end{enumerate}

The second step above corresponds to gradient ascent of $L_c(\bm{x},\bm{\lambda})$ in the $\bm{\lambda}$ variables. Basically, this method performs a \emph{fast gradient descent} of $L_c$ in the $\bm{x}$ directions in conjunction with a \emph{slow gradient ascent} of $L_c$ in the $\bm{\lambda}$ directions. A dynamical system that performs this process in continuous time is given below:
\begin{align*}
    \frac{dx_i}{dt}&=-\kappa\frac{\partial}{\partial x_i}L_c(\bm{x},\bm{\lambda}),\\
    \frac{d\lambda_i}{dt}&=\kappa'\frac{\partial}{\partial \lambda_i}L_c(\bm{x},\bm{\lambda}).
\end{align*}

\section{Application of Lagrange multipliers to the Ising problem; Cubic terms}
In this appendix, we shall provide an explanation of the role that is played by the cubic terms---that were neglected in the main text---in the methods of Yamamoto et al. \cite{haribara_computational_2016}, Kalinin et al. \cite{kalinin_global_2018}, and others. It will turn out that the inclusion of cubic terms helps us implement the `Augmented Lagrangian method of multipliers' from Appendix B. 

The statement of the Ising problem is as follows:
\begin{alignat*}{2}
    &\text{minimize}\ \ \ &&\bm{x}^T\bm{J}\bm{x}\\
    &\text{subject to }\ \ &&x_i^2-1=0,\ \ i=1,\dots,n.
\end{alignat*}
The corresponding Lagrange function is given by:
\begin{equation*}
    L(\bm{x},\bm{\lambda})=\bm{x}^T\bm{Jx}-\sum_{i=1}^n\lambda_i(x_i^2-1).
\end{equation*}
Next, we write down the Augmented Lagrange function:
\begin{equation*}
    L_c(\bm{x},\bm{\lambda})= \bm{x}^T\bm{Jx}-\sum_{i=1}^n\lambda_i(x_i^2-1)+\frac{c}{2}\left(\sum_{i=1}^n\left(x_i^2-1\right)^2\right).
\end{equation*}
Substituting the above Augmented Lagrange function into the dynamical system provided at the end of Appendix B, we get:
\begin{align*}
    \frac{dx_i}{dt}&=2\kappa\left(\lambda_ix_i+2cx_i-2cx_i^3-\sum_{j=1}^n{J_{ij}x_j}\right),\\
    \frac{d\lambda_i}{dt}&=\kappa' (1-x_i^2),
\end{align*}
where $\kappa$ and $\kappa'$ are appropriately chosen step sizes. We notice the equivalence in form between this dynamical system and those in the papers discussed in the main text and conclude that the cubic terms that appear in most of those systems are in fact helping to implement the Augmented Lagrangian method of multipliers. 

\section{Adiabatic method using Kerr $\lp\chi^{(3)}\rp$ nonlinear coupled oscillators}
Researchers at Toshiba proposed an adiabatic Ising solver consisting of networks of coupled Kerr nonlinear oscillators \cite{goto_combinatorial_2019}, \cite{goto_bifurcation-based_2016}, \cite{goto_quantum_2019}. The big difference between the machines we have studied so far and the Toshiba machine is that Toshiba uses dissipation-less optical systems and utilizes the adiabatic method to optimize the Ising Hamiltonian. By dissipation-less, we mean that the coupling between different oscillators is not dissipative but perfectly elastic.

Further, they replace parametric oscillators that use the $\chi^{(2)}$ nonlinearity with Kerr-parametric oscillators that possess both the $\chi^{(2)}$ and the $\chi^{(3)}$ nonlinearities. The $\chi^{(2)}$ nonlinearity gives rise to effects such as parametric amplification whereas the $\chi^{(3)}$ nonlinearity leads to intensity dependent refractive index change. A second-harmonic pump signal is used to achieve parametric amplification. The refractive index of the modes in the oscillators is modulated by their own intensities due to the third-order nonlinearity.

The slowly varying amplitude equations for the sine ($s_i$) and cosine ($c_i$) components of the electric field in their system are given by:
\begin{gather*}
    \frac{dc_i}{dt}=K\lp c_i^2+s_i^2\rp s_i+ps_i+\xi_0\sum_{j}{J_{ij}s_j},\\
    \frac{ds_i}{dt}=-K\lp c_i^2+s_i^2\rp c_i+pc_i-\xi_0\sum_{j}{J_{ij}c_j},
\end{gather*}
where $c_i$ and $s_i$ represent the cosine and sine components of the $i$-th oscillator, $p$ refers to the strength of the parametric pumping, $K$ is the value of the third-order nonlinearity (Kerr coefficient), and $\xi_0$ is the strength of the coupling interaction between the oscillators.

We rotate the axes and recast the above set of equations in phasor notation as follows to elucidate the similarity between this method and that of Cambridge:
\begin{equation*}
    \frac{dE_i}{dt}=\kappa\lp pE_i^*+iK|E_i|^2E_i+i\sum_j{J_{ij}E_j}\rp.
\end{equation*}
The first term on the right-hand side is the parametric gain, the second term is the nonlinear rotation caused by the Kerr nonlinearity, and the third term is the dissipation-less coupling between oscillators.

The machine, which follows the adiabatic principle, works by ramping up the value of the pump $p$ adiabatically. The authors show that their system Hamiltonian mimics the Ising Hamiltonian well when p becomes large enough.

We note the interesting fact that, though the authors have parametric gain, the system does not blow up to infinity due to the presence of the nonlinear rotating term. This term ensures that the quadrature vector keeps traversing gain and loss regions periodically in the phase space, hence keeping the amplitudes in check.\\

\section{Iterative Analog Matrix Multipliers}
In this appendix, we present the system designed by Soljacic et al. \cite{roques-carmes_heuristic_2020}. We shall see that the simplified system we presented in Sec 6a of the main text differs from that of Soljacic et al. significantly in that their method has added noise and nonlinear thresholding after each iteration. It is possible that their modifications lead to performance improvements.

Formally, their iteration is given by:
\begin{equation*}
    \bm{E}(t+1)=u\lp2\bm{KE}(t)+\bm{N}(t)\rp,
\end{equation*}
where $\bm{E}(t)$ is the vector of electric field amplitude values at the beginning of the $t$-th iteration, $u(x)$ is the Heaviside step function that is 1 for positive $x$ and 0 for negative $x$, $\bm{N}(t)$ is a zero-mean Gaussian random noise vector, and $K$ is a matrix given by $K=\sqrt{\bm{J}+\alpha \bm{M}}$, $\bm{J}$ is the Ising connectivity matrix, $\alpha$, a real number, and $\bm{M}$, some suitably chosen matrix. More specifically, $\bm{M}$ is chosen to have the same eigenvectors as $\bm{J}$. It will turn out that the eigenvalues of $\bm{M}$ play the role of Lagrange multipliers.

The authors showed that under the condition of high noise $\bm{N}(t)$, their system performs minimization of the following effective merit function:
\begin{equation*}
    H=-\frac{\beta}{2}\sum_{ij}{\lp J_{ij}+\alpha M_{ij}\rp E_iE_j},
\end{equation*}
where $\beta$ is some parameter dependent on the noise.

Using the fact that the matrix $\bm{M}$ is chosen to have the same eigenvectors as $\bm{J}$, we rewrite the above merit function, modulo additive constants, as the following Lagrange function:
\begin{equation*}
    H=L(\bm{E},\bm{\gamma})=-\frac{\beta}{2}\lp\sum_{ij}{J_{ij}E_iE_j}+\alpha\sum_i{\gamma_i\lp z_i^2-1\rp}\rp,
\end{equation*}
where the $\gamma_i$ are the eigenvalues of the matrix $\bm{M}$, and the vector $\bm{z}$ is the vector of electric field amplitudes $\bm{E}$ expressed in the basis of eigenvectors of $\bm{M}$ (that is, the eigenvectors of $\bm{J}$). We see that the eigenvalues of $\bm{M}$ play the role of Lagrange multipliers, albeit for different constraints than those required by the Ising problem. This difference seems to be caused by $\bm{M}$ not being a diagonal matrix.

In conclusion, we interpret their algorithm as optimizing a Lagrange function with the merit function being the Ising Hamiltonian itself, and the constraints being that the components of the spin vector when expressed in the eigenvector basis of $\bm{J}$ be restricted to 1 and -1.

\bibliography{Reffinal}
\end{document}